\documentclass[aip]{revtex4-1}
\usepackage{xcolor}
\usepackage{graphicx}% Include figure files
\usepackage{dcolumn}% Align table columns on decimal point
\usepackage{bm}% bold math

\begin{document}

%\preprint{}

\title{Point process analysis of geographical diffusion of news in Argentina} %Title of paper



\author{L. L. García}
\email[]{llgarcia@df.uba.ar}
%\thanks{}
\affiliation{Departamento de Física, Universidad de Buenos Aires, Facultad de Ciencias Exactas y Naturales, 1428, Buenos Aires, Argentina}
\affiliation{Instituto de Física Interdisciplinaria y Aplicada (INFINA), CONICET - Universidad de Buenos Aires, Buenos Aires, Argentina}

\author{G. Tirabassi}
\email[]{giulio.tirabassi@upc.edu}
%\thanks{}
\affiliation{Departament de Física, Universitat Politécnica de Catalunya, Rambla St. Nebridi 22, Terrassa, 08222, Barcelona, Spain}
\affiliation{Departament de Informàtica, Matemàtica Aplicada i Estadística, Universitat de Girona, Carrer de la Universitat de Girona 6, Girona, 17003, Spain}

\author{C. Masoller}
\email[]{cristina.masoller@upc.edu}
%\thanks{}
\affiliation{Departament de Física, Universitat Politécnica de Catalunya, Rambla St. Nebridi 22, Terrassa, 08222, Barcelona, Spain}

\author{P. Balenzuela}
\email[]{balen@df.uba.ar}
\affiliation{Departamento de Física, Universidad de Buenos Aires, Facultad de Ciencias Exactas y Naturales, 1428, Buenos Aires, Argentina}
\affiliation{Instituto de Física Interdisciplinaria y Aplicada (INFINA), CONICET - Universidad de Buenos Aires, Buenos Aires, Argentina}


\date{\today}

\begin{abstract}

The diffusion of information plays a crucial role in a society, affecting its economy and the well-being of the population. Characterizing the diffusion process is challenging because it is highly non-stationary and varies with the media type. To understand the spreading of newspaper news in Argentina, we collected data from more than 27000 articles published in six main provinces during four months. 
We classified the articles into 20 thematic axes and obtained a set of time series that capture daily newspaper attention on different topics in different provinces. To analyze the data we use a point process approach. For each topic, $n$, and for all pairs of provinces, $i$ and $j$, we use two measures to quantify the synchronicity of the events, $Q_s(i,j)$, which quantifies the number of events that occur almost simultaneously in $i$ and $j$, and $Q_a(i,j)$, which quantifies the direction of news spreading. Our analysis unveils how fast the information diffusion process is, showing pairs of provinces with very similar and almost simultaneous temporal variations of media attention. On the other hand, we also calculate other measures computed from the raw time series, such as Granger Causality and Transfer Entropy, which do not perform well in this context because they often return opposite directions of information transfer. We interpret this as due to different factors such as the characteristics of the data, which is highly non-stationary and the features of the information diffusion process, which is very fast and probably acts at a sub-resolution time scale.

\end{abstract}


\maketitle %\maketitle must follow title, authors, abstract and \pacs

\begin{quotation}
Throughout his distinguished scientific career, Professor Jason Gallas tackled a wide range of challenges related to complex systems and nonlinear dynamics, covering both fundamental and applied areas. His research encompassed diverse topics, such as economic models where he examined the simultaneous variation of multiple parameters in economic processes, social dynamic models focused on stress-related behaviors, and the analysis of science communication statistics, particularly citation patterns among physicists, to explore whether the prevalence of mathematical equations in a paper influences its citation count. In this contribution to the Focus Issue in Honor of Professor Gallas, we study news communication and public opinion dynamics in Argentina. We analyze a large number of newspaper articles published in six main provinces during four months. We find very similar and almost simultaneous temporal evolution of media attention in different provinces, which can be interpreted as due to fast information diffusion produced by social networks (Instagram, X, etc.) and the action of large media companies that act as global, external drivers of information diffusion.
\end{quotation}

\section{Introduction}\label{Introduccion}

During his prolific scientific career, Professor Jason Gallas addressed a wide variety of problems involving complex systems and nonlinear dynamics, of fundamental and applied interest, including topics as diverse as economic models, focusing on the simultaneous variation of several parameters involved in economic processes \cite{Gallas_96}, social dynamic models of people's stress-related processes \cite{Gallas_17}, and science communication statistics, analyzing citation habits among physicists, to determine whether mathematics (i.e., the density of equations in a paper) can be blamed for the number of citations \cite{Gallas_15}. In this contribution to the Focus Issue in Honor of Professor Gallas, we study news communication and public opinion dynamics in Argentina.

The mass media play a preponderant role in the formation of public opinion. The systematic study of this dynamics is known as {\it{agenda-setting theory}} and began with the seminal work of McCombs and Shaw in 1972 \cite{mccombs1972}. The agenda-setting theory compares the public agenda with the issues covered by the media, i.e., the issues that the public considers a priority, and the relationship between public and media perceptions of political figures, celebrities, countries, conflicts, events, etc. \cite{Tsur_2015, 16_Paper_Seba, Cosenza2020}. 

The agenda-setting theory studies how the media shapes public opinion by giving particular attention to some topics. In the last few years, access to big data has enabled modeling and  quantitative analyses, such as studies of correlation between media and public attention, information and opinion spreading \cite{Cozzo_2017, Kuperman_2005, MGE_2023, Nuno_2012, 15_Paper_Seba, 2_Paper_Seba, 16_Paper_Seba, 17_Paper_Seba, 18_Paper_Seba}. Recently, an unsupervised topic decomposition model on newspaper articles was developed to study the dynamics of mass media and public agendas \cite{PaperSeba}. In \cite{albanese2020analyzing} the topic decomposition model, combined with sentiment analysis and time series analysis methods, was used to analyze news articles during the 2016 US presidential campaign, to obtain insights into the relationship between media and public opinion, using voting intention as a proxy of public opinion.

The study of how different news media influence each other, in determining what issues to cover and how they present the news to the public, is now known as {\it{intermedia agenda-setting}} \cite{McCombs_2004}. Studies on this line of research typically indicate that the coverage often shifts from elite news sources to other media platforms \cite{Reese&Danielian_1989, Golan_2006, Kushin_2010}. Guo and Vargo \cite{Guo&Vargo_2017} analyzed how news media in different countries influence each other in covering international news. Mohammed and McCombs \cite{McCombs_2021} studied the flow of international news and provided evidence that media such as {\it The New York Times} or {\it The Guardian} do not publish much news about peripheral countries, and when they do, it is mostly negative news. Guo and Zhang \cite{Guo_2023} analyzed the information flow in the United States, examining data from the 21 most populated US cities, and characterized how national news and local news coverage influence each other. Stern and collaborators \cite{Stern2020} used a network perspective to study networks of influence between different news sources on particular topics, and found that the networks depend significantly on the topic, since some news sources may act as agenda-setters (i.e., central nodes) in certain topics while they may act as followers (i.e., peripheral nodes) in others topics. Another example of information diffusion modeling in real networks is the one carried out  in \cite{WCota_2019}. There, they construct a political communication network on Twitter and analyze the diffusion of information among users using epidemiological models.

The flow of information between different regions or countries has also been studied using complex network representations. Examples, to name just a few, are the analysis of millions of geo-coded external hyperlinks, collected from Danish, Norwegian, and Swedish news websites \cite{2018_Scandinavian}; the analysis of millions of news articles used to construct a network based on the sequential mention of events across various countries \cite{2024_Quattrociocchi}, and the study of geolocalized time series
of tweets associated to five events (the release of a Hollywood movie, two political protests in Brazil and in Spain, the Higgs boson discovery, and the acquisition of Motorola by Google), to extract, using transfer entropy analysis, directed networks that capture the role played by different geographical units in the global exchange of information \cite{2016_Vespingnani}. 

In this work, we characterize the diffusion of information in newspapers in Argentina, by analyzing more than 27000 news articles published in six main provinces during four months. We selected a time range that included two events that attracted a lot of press attention: the arrival in Argentina of an Iranian Venezuelan plane on June 6, 2022, and the attempt to assassinate the Vice President of Argentina, on September 1, 2022.
We are interested in characterizing how specific news diffuse and gain broad national attention. We extract the media agenda by decomposing the topics in 20 non-orthogonal topics, using the unsupervised topic decomposition model developed in \cite{PaperSeba}. Then, we use a “point process”, event-based approach \cite{quianquiroga2002} to detect the onset of coverage of different topics in different provinces. This approach is based on the hypothesis that a sudden increase in media coverage is triggered by a well-defined and highly informative event. We characterize the synchronicity of media attention events in pairs of provinces, $i$ and $j$ (the events are defined as threshold-crossings of the attention on a topic in a province) and the direction of news spreading (if attention first starts in province $i$, or in province $j$). We also analyze the interdependence of news coverage between different provinces by using well-known bivariate measures to detect linear and nonlinear correlations and dependencies. 

The paper is organized as follows: Sec.~\ref{Metodos} presents the data analyzed and the methods used, Sec.~\ref{Resultados} presents the results obtained, and Sec.~\ref{Conclusiones} presents the discussion of the results and outlook.

% If in two-column mode, this environment will change to single-column format so that long equations can be displayed. 
%\begin{widetext}
%$$\mbox{put long equation here}$$
%\end{widetext}

\section{Data and Methods}\label{Metodos}

\subsection{Data}

The database analyzed consists of 27,881 political news articles collected from news papers published in the six Argentinian provinces listed in Table \ref{tab:database}. 
The study period spans from May 26, 2022, to September 26, 2022. For data collection, the two main media outlets in each region were selected. 
The sources and the number of collected articles are listed in Table \ref{tab:database}.  

% \begin{table}
% \end{tabular}
% \end{table}

\begin{table}[htpb]
\caption{\label{tab:database} Number of articles of each province and the hyperlinks to the news media where they were published.}
%\centering
\begin{tabular}{|c|c|c|c|}
\hline
\textbf{Number} & \textbf{Province} & \textbf{No. of articles} & \textbf{Media} \\
\hline
1 & Buenos Aires  & 6415  & \href{https://www.clarin.com/}{Clarín},  \href{https://www.lanacion.com.ar/}{La Nación} \\
\hline
2 & Córdoba & 5013  & 
\href{https://www.lavoz.com.ar/}{La Voz},  \href{https://www.cadena3.com/}{Cadena 3} \\
\hline
3 & Tucumán & 5552 & 
\href{https://www.lagaceta.com.ar/}{La Gaceta},  \href{https://www.losprimeros.tv/}{Los Primeros}  \\
\hline
4 & Mendoza & 6359 & 
\href{https://www.losandes.com.ar/}{Los Andes},  \href{https://www.mdzol.com/}{MDZ Online}  \\
\hline
5 & Santa Fe & 2037  & 
\href{https://www.ellitoral.com/}{El Litoral},  \href{https://www.lacapital.com.ar/}{La Capital} \\
\hline
6 & Santiago del Estero & 2505 & 
 \href{https://www.diariopanorama.com/}{Diario Panorama},  \href{https://www.nuevodiarioweb.com.ar/}{Nuevo Diario} 
 \\
\hline
\end{tabular}
\end{table}

\subsection{Topic decomposition model}  
\label{metodo_agenda}

Here we describe how the content of the articles was analyzed and the methodology used for topic decomposition.
    
\textbf{Pre-Processing Text:} Firstly, we performed text tokenization by removing all the accents and by removing stop-words from a custom vocabulary which contained articles, verbs, and adverbs. In addition, very frequent words (present in more than 95$\%$ of the articles) and very rare words (present at most in 2 articles) were discarded.

\textbf{Topic Decomposition:} The texts were represented as vectors in a mathematical space using the term frequency - inverse document frequency (tf-idf) representation \cite{nltk}. Each element of this vector represents words based on their weighted frequency by specificity \cite{NGUYEN201495}. A corpus composed by a group of news articles is then represented as a matrix, $M$, where rows correspond to articles, and columns represent the words used to describe them. Finally, from this matrix, a number of topics, $N$, were identified using the non-negative matrix factorization (NMF) algorithm from the Python library scikit-learn \cite{sklearn} that approximates $M$ as the product of two matrices, $H$ and $W$, that represent documents in the topic space and topics in the word space, respectively. The number of topics $N$ is a parameter of the algorithm. The algorithm ensures that all the elements of $H$ and $W$ are non-negative, which results in {\it non-orthogonal topics}. 
By normalizing the rows of $H$ to the L1 norm, the elements of each row ($h$) can be interpreted as the degree of membership of a document ($k$) to the n-th topic, while the rows of $W$ detail the characterization of each topic using the vocabulary ($w_{nj}$ indicates the weight of term $j$ in topic $n$, meaning how effectively term $j$ encapsulates the content of topic $n$). 

\textbf{Description of the topics:} Finding the optimal number of topics, $N$, is always a challenge in topic models. After trying several values, we decided to select $N=20$ because it provided a set of topics with clear interpretation (as it can be seen in the description below) . Selecting a larger number of topics can reduce the topic-overlap, but can also introduce “small topics” that are not really relevant in the global agenda (but that may be locally important in different Argentinian provinces), while selecting a lower number of topics will merge different topics, making the topics more generic. Clearly, the number of topics (as well as the event-detection thresholds and link-detection thresholds to be described in the next subsections), are hyper-parameters that have to be setup heuristically. Additional studies are planed to clarify the role of these parameters.

The topics are described as: 

\begin{enumerate}
    \item \textbf{Assault on Vice President}: The attempt of assassination of Vice President Cristina Fernandez by Fernando Sabag Montiel. 
    \item \textbf{Vice President}: Includes press articles that mention the vice president, including articles that mention the former president, as well as articles that mention the assassination attempt. Topic 1 and topic 2 have a large overlap because, as explained before, the topic decomposition is non-orthogonal.
    \item \textbf{Diplomatic conflict}: Includes press articles that mention  the arrival of a plane from Venezuela with Iranian crew. 
    \item \textbf{Former Economy Minister}: Includes press articles that mention the former economy minister, Silvina Batakis; the analyzed time period covers Batakis' resignation. 
    \item \textbf{Economy Minister}: Includes press articles that mention the new economy minister, Sergio Massa. 
    \item \textbf{Gas Pipeline}: Includes press articles that mention the construction of the gas pipeline Nestor Kirchner. 
    \item \textbf{Social Leader}: Includes press articles that mention the health condition of social leader Milagro Sala, from {the province of} Jujuy (which is not one of the six main provinces included in this study). 
    \item \textbf{Domestic politics}: Includes press articles that mention the former governor of Tucumán, Juan Manzur, and his assumption as chief of staff. 
    \item \textbf{Vice President Trials}: Includes press articles that mention several legal cases of Vice President, Cristina Kirchner, particularly the “Causa Vialidad”. 
    \item \textbf{Dollar}: Includes press articles that mention the value of the dollar. 
    \item \textbf{Supreme Court}: Includes press articles that mention the Supreme Court and the proposed bill to modify it. 
    \item \textbf{Government}: Includes press articles that mention President Alberto Fernandez. 
    \item \textbf{Inflation}: Includes press articles that mention the inflation rate. 
    \item \textbf{Fuel Shortage}: Includes press articles that mention the shortage of diesel fuel in different regions. 
    \item \textbf{Strikes}: Includes press articles that mention the strikes carried out by the truckers' union, partly due to the Fuel Shortage. 
    \item \textbf{Libertarian Party}: Includes press articles that mention Javier Milei, the main politician of the libertarian party in Argentina (actual president of Argentina).
    \item \textbf{Subsidy Segmentation}: Includes press articles that mention the removal of energy subsidies in Buenos Aires and its segmentation. 
    \item \textbf{Opposition}: Includes press articles that mention actions and sayings of the coalition of opposition parties to the government of Alberto Fernández.
    \item \textbf{International affairs}: Includes press articles that mention international politics, mainly the “Cumbre de las Américas”.
    \item \textbf{Social Assistance}: Includes press articles that mention social assistance and the social organizations.
\end{enumerate}

Additional information can be found in  Appendix \ref{secA1}: the word-clouds of the topics are shown in Fig.~\ref{fig_wordclouds}, and the five most relevant keywords of each topic are listed in Table \ref{tab:topics}.

\subsection{Media agenda}  
Here we define the global and local media agenda, which are time series that quantify the attention received by the topics described before, as a function of time, at the national level, and at a local level, in the six main provinces listed in Table \ref{tab:database}.

If the topics were mutually orthogonal, the agenda could simply be calculated as the number or news articles belonging to each topic. Since they are not, the agenda is built by adding the degree of membership (given by the elements of matrix $H$, $h_{k,n}$) to each topic $n$, of each article, $k$, published in a period of time, $[t, t+\Delta t]$, normalized to the total number of articles, $L$, published in the same time interval, 
\begin{equation}
 x^n(t) = \frac{1}{L} \sum_{k \in [t,t+\Delta t] } h_{k,n}
 \label{eq:global_definition}
\end{equation}
In this way, we obtained a set of $N$ time series that we refer to as {\it global} media agenda, which quantify the total coverage received by the $N$ topics. The data values are normalized such that each day, adding the media attention over all the topics gives 1, $\sum_n x^n(t)=1$ for all $t$. 

To obtain the {\it local} agenda of topic $n$ in province $i$ we added the elements of the matrix $H$, $h_{k,n}$, over all the documents, $k$, that were published within $[t, t+\Delta t]$, in a media outlet located in the province $i$, normalized to the total number of articles $L_i$ published in the same time interval. 

\begin{equation}
 x^n_i(t) = \frac{1}{L_{i}} \sum_{k \in [t,t+\Delta t], k \in i } h_{k,n}
 \label{eq:topic_definition}
\end{equation}
In this way, the data values are normalized such that in each province, each day, the media attention over all the topics is 1, that is $\sum_n x^n_i(t)=1$ for all $i$, for all $t$.

\begin{figure}[tb!]
\centering
\includegraphics[width=0.9\textwidth]{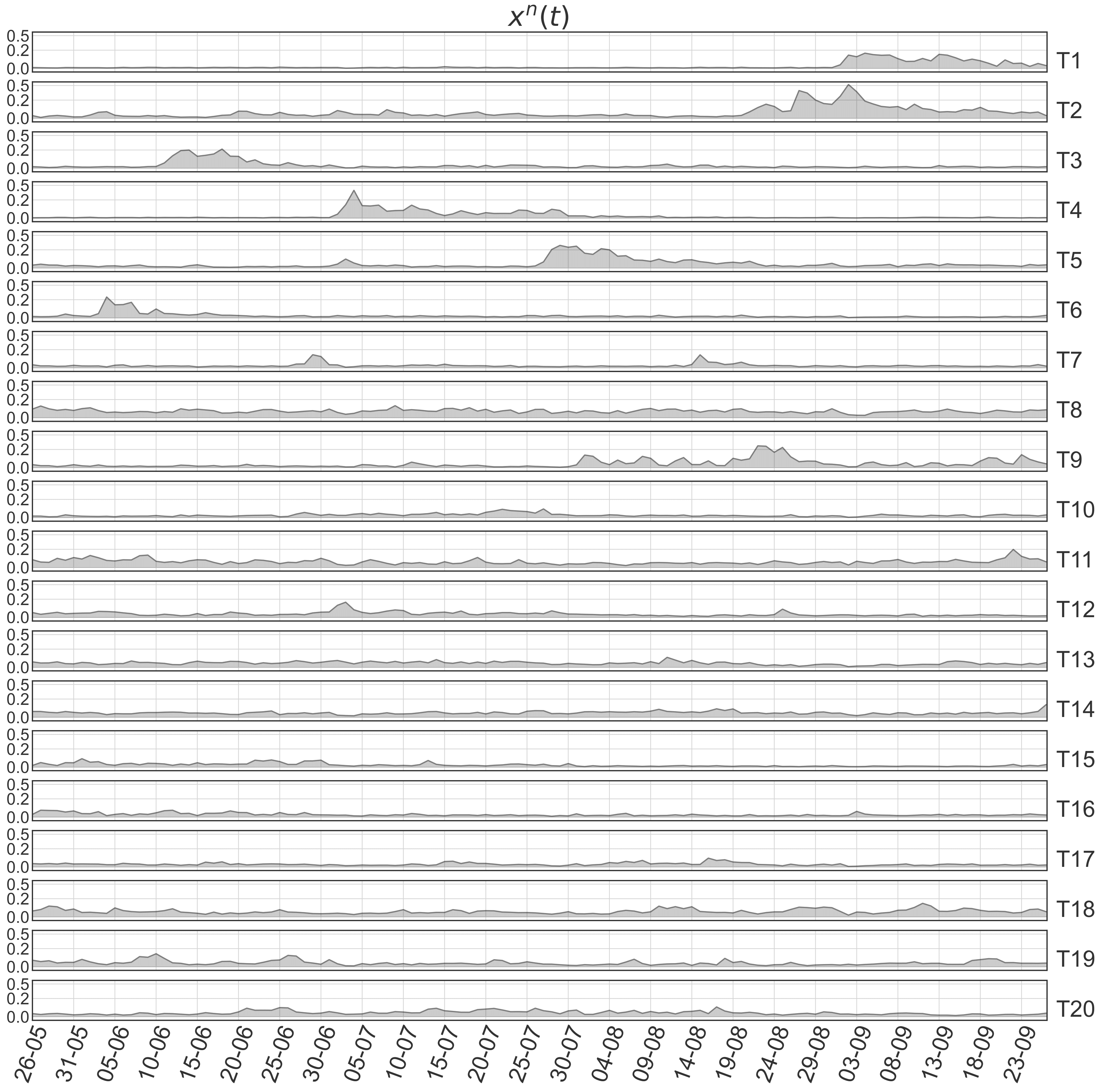}
\caption{ \label{fig_global_agenda} \textbf{Global agendas} of the 20 topics. The notation $T_{i}$ denotes the i-th topic following the labels assigned in \ref{tab:topics}. Some topics have time series with pronounced events while other do not. The topics are numbered and displayed from more (top) to less (bottom) eventful. }
\end{figure}

Figure \ref{fig_global_agenda} shows the time series of global media agenda obtained for the 20 topics described before. It can be observed that there are topics with well-defined peaks, usually triggered by particular events, such as “Vice President” or “Diplomatic conflict” or “Social Leader”, while others remain on the agenda consistently over time, contributing to the agenda as background topics, such as “Dollar”, “Inflation” or “Government”. 

It is important to remark that a noisy “background” can be just a numerical artifact. In fact, depending on the topic, no media attention was paid to it before a particular event actually occurred. A clear example of this situation are the noisy values of the agenda time series of topic $n=3$ (Diplomatic conflict) before the plane that triggered the conflict arrived in Argentina on June 6, 2022. Therefore, all the values of the time series previous to June 6 should be considered spurious. Another example are the agenda values of topic $n=1$ (Assault on Vice President) before the assault occurred on September 1, 2022. This motivate the “event-based” point process method used to analyze the data and it may also partially explain the bad performance of well-known methods to quantify information transfer from raw time series analysis, Granger causality and transfer entropy, as will be discussed latter.

\begin{figure}[tbp!]
\centering
\includegraphics[width=\textwidth]{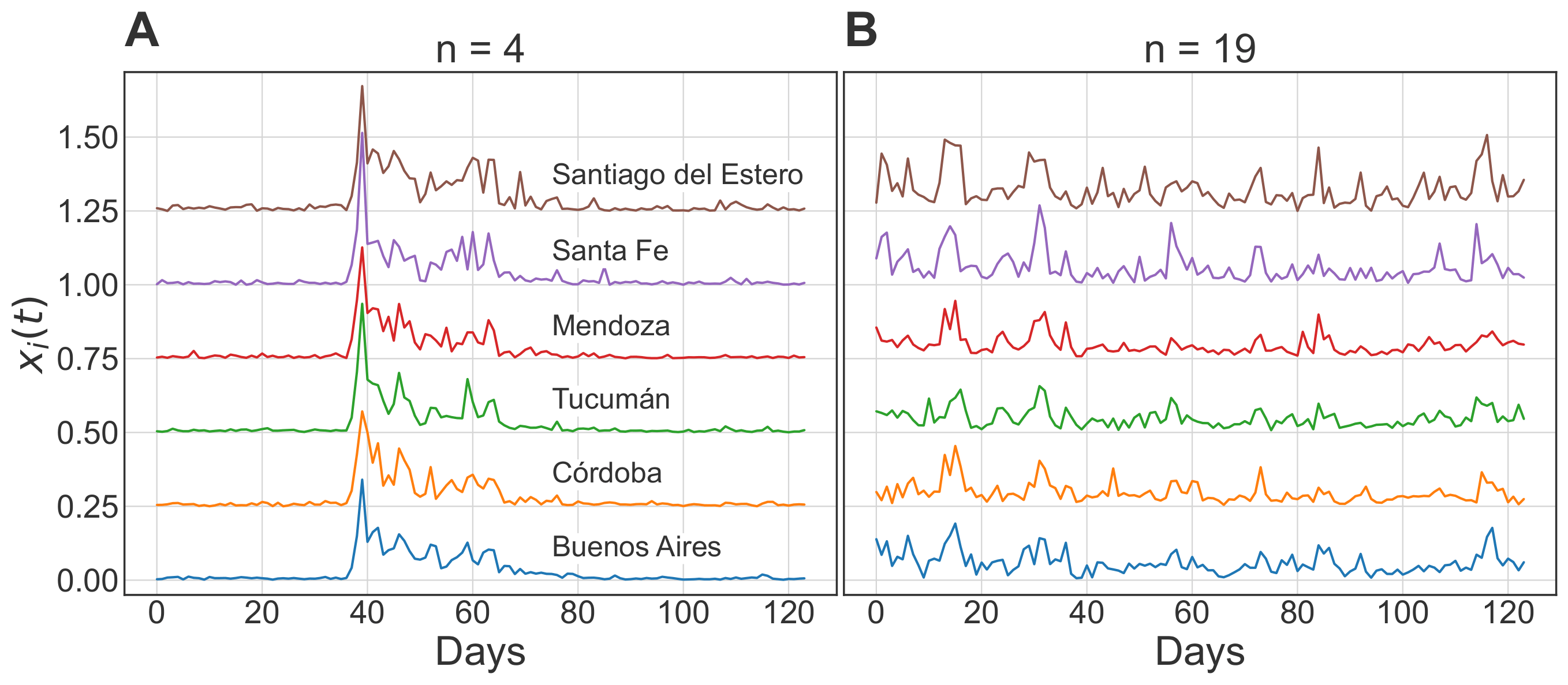}
\caption{\label{fig_agenda_provinces} \textbf{Local agendas} of two topics in the six provinces (shifted vertically for clarity). \textbf{A} Topic “Former Minister of Economy”; here we note that all time series exhibit a well-defined event that subsequently decays. \textbf{B} Topic “International affairs”; here we see a much noisier behavior, with the time series displaying numerous peaks. }
\end{figure}

Two examples of the local agendas are displayed in Fig.~\ref{fig_agenda_provinces}, where we see that when a topic is triggered by a particular event, the evolution is quite similar in the different provinces, while if the topic is consistently present in the agenda, its evolution can be quite different in the different provinces.
The values of the data points vary with the particular topic and province, because a topic in a province may not be as relevant as in another.

% Figures should be put into the text as floats. 
% Use the graphics or graphicx packages (distributed with LaTeX2e).
% Put the label that you will use with \ref{} command in the braces of the \label{} command.
%
% \begin{figure}
% \end{figure}

\subsubsection{Point process description of the local agendas}  
\label{seccion_metodo_discretizacion}

To define events of media attention in the local agendas, we used an activation threshold and a deactivation one. An event for the $n$-th topic in the $i$-th province is identified when the time series $x^n_i(t)$ crosses the activation threshold $th_{a}$. Then, before another event can occur, we required that the agenda value decreases below the deactivation threshold, $th_{d}<th_{a}$.

The thresholds were defined for each topic, in terms of the mean, $\mu_n$, and the standard deviation, $\sigma_n$, of the global agenda of the $n$-th topic. Specifically, the thresholds were defined as:
    \begin{eqnarray}
        th_{a}^{n} &=& \mu_n + \sigma_n \nonumber
        \\
        th_{d}^{n}  &=& \mu_n
     \label{eq:thresholds}
    \end{eqnarray}

\begin{figure}[tbp!]
    \centering
    \includegraphics[width=0.7\textwidth]{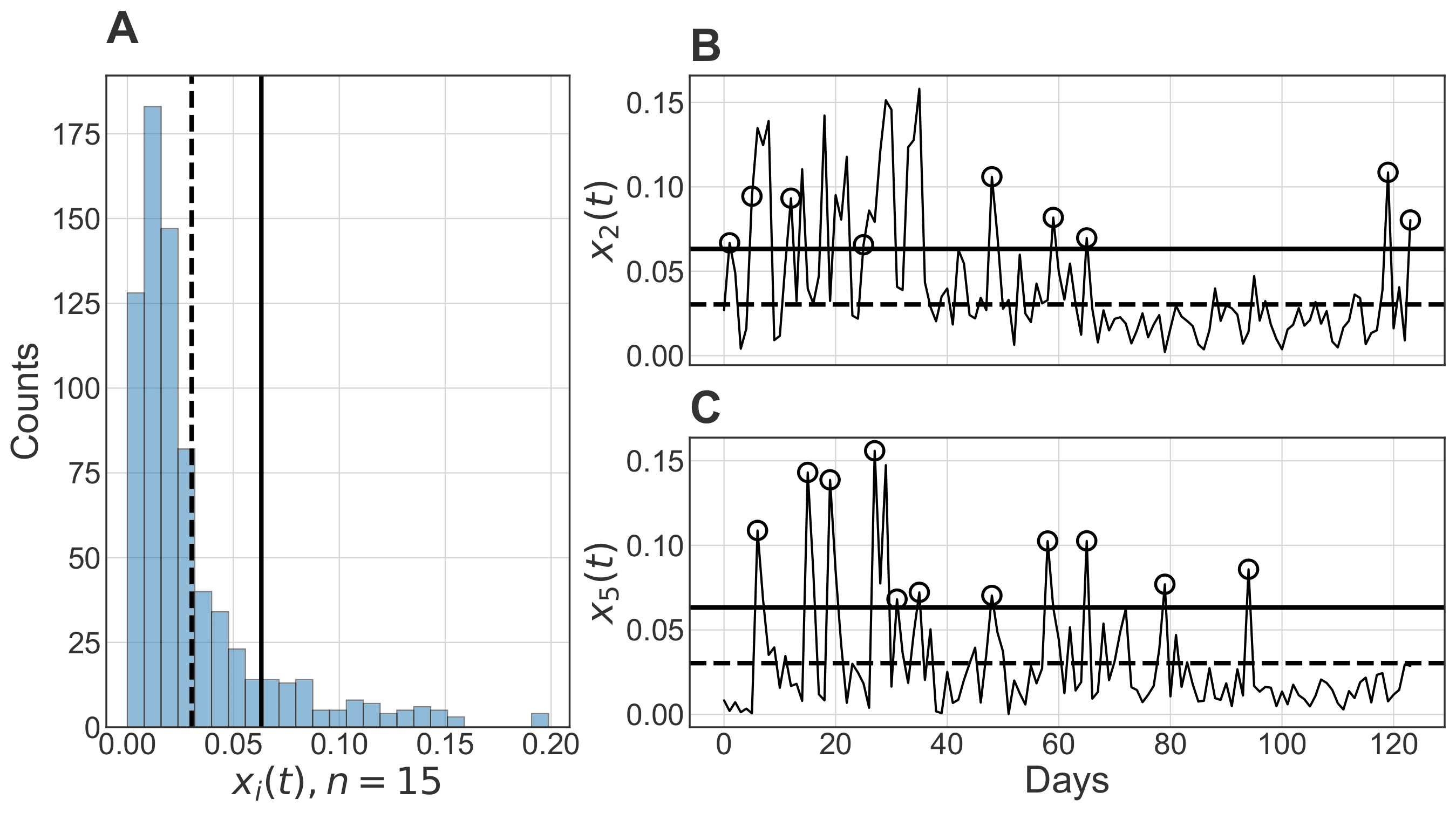}
    \caption{\label{fig_event_method} \textbf{A} Histogram of all the local agenda values (for the six provinces) of topic “Fuel shortage” ($n=15$). The vertical lines indicate the activation and deactivation thresholds, $th_a$ (solid line) and $th_d$ (dashed line). \textbf{B} \& \textbf{C}, Local agendas in Córdoba ($i=2$) and Santa Fe ($i=5$), respectively; the circles indicate the events detected.}
\end{figure}

Figure~\ref{fig_event_method} shows an example of the distribution of global agenda values of a topic, “Fuel shortage” ($n=15$). The figure also shows the activation and  deactivation thresholds of the topic, and the events detected in two provinces, Córdoba and Santa Fe.

The values selected for the two thresholds allow us to detect when media attention on a topic starts to grow, and also, as it will be shown later, they allow detecting events in all the time series. 

Obviously, our results will depend on the event-detection strategy and thresholds employed. We tested a different strategy, based on using a single threshold to detect an event --the onset of attention in a topic in a province-- and requiring that a short time interval passes, $\Delta t$ (of two-three days), before another event can occur. However, with this strategy we found that fast and large variations in the agenda (see, for example, Figs.~\ref{fig_event_method}\textbf{B}, Figs.~\ref{fig_event_method}\textbf{C}) were not detected as different media attention events. Therefore, we decided to use two thresholds, one to detect an increase of attention in a given topic, and the other, to detect when the attention falls (and then, another different event may occur latter). We also tested defining individual thresholds for each time series, in terms of the mean and standard deviation of the local agenda values, and found qualitatively similar results. Since already a simple definition of the thresholds, Eq.(\ref{eq:thresholds}), leads to a set of events which appear meaningful (particularly for topics with well-defined peaks), we keep this definition. Future work will be aimed at comparing the results obtained with different event-detection strategies: with or without temporal lag to separate different events, thresholds defined for each time series (for each topic, in each province), thresholds defined for each topic (used here) and global thresholds (same for all the topics and for all the provinces).

\subsubsection{Event coincidence detection}

We used the method proposed in \cite{quianquiroga2002} to quantify the synchronicity of events in two different time series. 
The method is based on counting the number of quasi-simultaneous occurrences of events,
allowing for a short delay, $\tau$, between them. Here we use $\tau = 3$ days because, after carefully inspecting the relation among articles, it is approximately the maximum amount of time it takes for a news to appear elsewhere. 
While the method is generic and can be applied to any pair of time series, here we use it to analyze pairs of agenda time series of the same topic in different provinces.

Therefore, if two events in the agendas of topic $n$ in provinces $i$ and $j$, $x^n_i$ and $x^n_j$, occur at times $t_i$ and $t_j$ with $|t_i-t_j| \le \tau$, we detect a coincidence. We define the quantity $C_{ij}$ as the number of coincidences for which $t_i<t_j$, and $C_{ji}$ as the number of coincidences for which $t_i>t_j$. If the two events occur simultaneously, the coincidence is shared and we add $1/2$ to $C_{ij}$ and 1/2 to $C_{ji}$. 

\textbf{Event synchronization measures:} With $C_{ij}$ and $C_{ji}$ we calculate the measures $Q_{s}$ (symmetric/synchronous) and $Q_{a}$ (asymmetric/asynchronous) proposed in \cite{quianquiroga2002}.

\begin{eqnarray}
    {Q_s}_{i,j}&=&\frac{2}{ m} \left[ C_{ij}+C_{ji} \right],
    \\
    {Q_a}_{i,j}&=&\frac{2}{ m} \left[C_{ij}-C_{ji} \right].   
\label{eq:syn_def}
\end{eqnarray}
Here, $m=m_i+m_j$, with $m_i$ and $m_j$ being the number of events detected in the $x^n_i(t)$ and $x^n_j(t)$.

With this definition, ${Q_s}_{i,j}= 1$ if, for each event in $x^n_i$ (in $x^n_j$), there is an event in $x^n_j$ (in $x^n_i$), and their timing is such that $|t_i-t_j| \le \tau$. In addition, ${Q_a}_{i,j} = 1$ (${Q_a}_{i,j}=-1$) if and only if the events in $x^n_j$ (in $x^n_i$) always precede those in $x^n_i$  (in $x^n_j$). Clearly, ${Q_s}_{i,j}={Q_s}_{j,i}$ and ${Q_a}_{i,j}=-{Q_a}_{j,i}$. 

\begin{figure}[tbp!]
    \centering
    \includegraphics[width=0.9\textwidth]{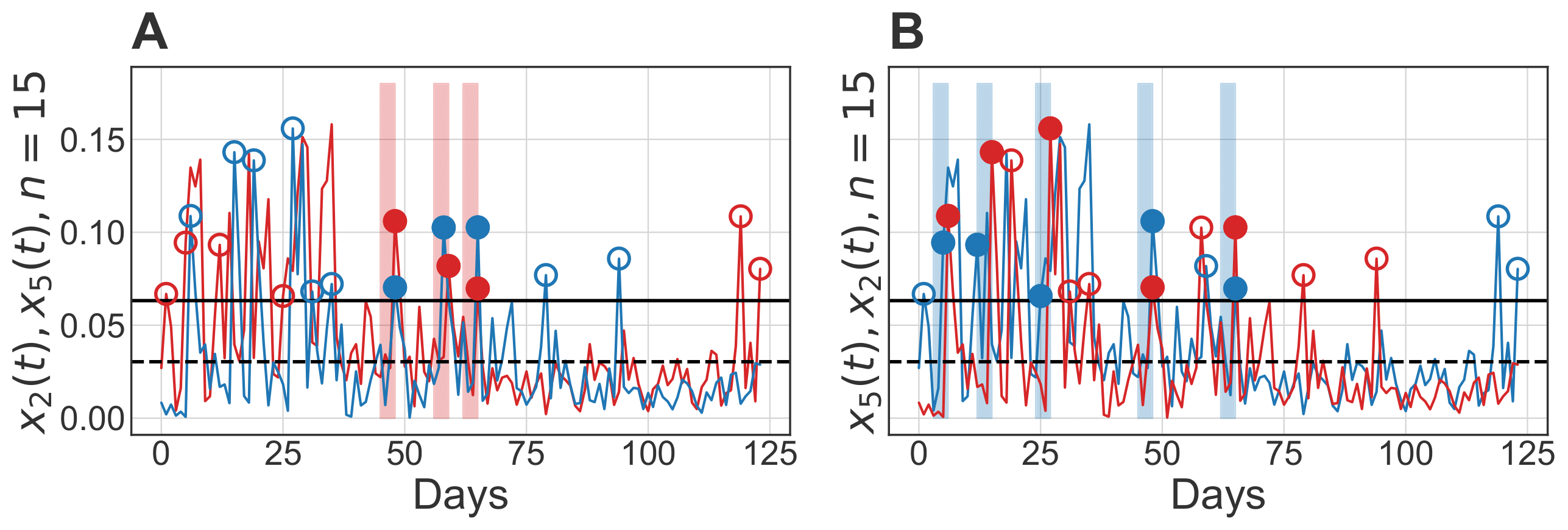}
    \caption{\label{fig_coincidence_method} \textbf{Event coincidence detection and quantification.} \textbf{A} Agenda time series of topic “Fuel Shortage” ($n=15$) in Córdoba ($i=2$, blue) and in Santa Fe ($j=5$ red); here colored circles indicate events in Córdoba (solid blue) that precede (by a maximum of $\tau = 3$ days) or that occur simultaneously to events in Santa Fe (solid red) --empty circles indicate all the other events.  Counting the number of coincidences gives $C_{2,5}(15) = 1/2+1+1/2=2$.
    \textbf{B} Time series “Fuel Shortage” in Santa Fe (blue) and in Córdoba (red). The colored circles show coincidences such that events in Santa Fe (solid blue) precede or occur simultaneously to events in Córdoba (solid red). Counting the number of coincidences gives $C_{5,2}(15) = 1+1+1+1/2+1/2=4$. Using Eq. \ref{eq:syn_def} with $m=20$ (total number of events) gives ${Q^{15}_s}_{5,2}=0.6$ and ${Q^{15}_a}_{5,2}=0.2$.}
\end{figure}

Figure \ref{fig_coincidence_method} illustrates how event coincidence detection works and how $Q_s$ and $Q_a$ are computed.

Summarizing, for each topic, $n$, from the events detected in the provincial agendas, we calculated the $6\times 6$ matrices ${Q^n_s}_{i,j}$ and ${Q^n_a}_{i,j}$, which are symmetric and anti-symmetric, respectively.

\subsubsection{Other analysis techniques used}

We also analyzed the local agendas using well-known bivariate measures to detect correlations and dependencies, which are computed from the raw time series: two undirected measures, linear cross-correlation, $CC$, and nonlinear mutual information, $MI$, and two directed measures, linear Granger causality, $GC$~\cite{granger1969}, and nonlinear Transfer entropy, $TE$~\cite{schreiber2000}. 
$GC$ starts from the assumption that the timeseries can be modeled using autoregressive processes with white noise, which implies timeseries with a Gaussian distribution. $MI$ and $TE$ do not require the distributions to be normal, however, it has been shown that the $k$-neighbours algorithm used to compute them works better if the timeseries are Gaussian-like and with similar variance \cite{kraskov2004estimating}. Therefore, we pre-processed the data by applying a log transformation to it, then we removed the mean to each timeseries and re-scaled them to unit variance. 

The linear cross-correlation was calculated using Pearson's equation,

\begin{equation}
CC(x,y) =\frac{\sum (x - \mu_x) (y - \mu_y)}{\sqrt{\sum (x - \mu_x)^2 \sum (y - \mu_y)^2}}.
\end{equation}

We calculated $CC$ using different time lags between the time series, but in most cases the maximum was found at zero lag, therefore, we decided to use zero-lag $CC$ as a synchronization metric.

MI is also a symmetric magnitude, that measures the amount of shared information between two time series, $x$ and $y$,
\begin{equation}
    MI(x, y) = \int\int p(x, y) \log\left(\frac{p(x, y)}{p_x(x) p_y(y)}\right) \mathrm{d}x\mathrm{d}y,
\end{equation}
where $p_x$ ($p_y$) is the probability density function of $x$ ($y$), and $p$ is the joint probability density function of $(x, y)$. To estimate the integral we used a $k$-neighbors algorithm with $k=4$, providing a good compromise between bias and variance \cite{kraskov2004estimating}.  

Measuring GC consists in modeling one time series as an autoregressive process of order $D$ and determining whether adding lagged values of another series improves the prediction of the autoregressive model, suggesting a potential causal relationship between them. 
In particular, to test $y\rightarrow x$ causality in Granger sense, we first fit the model 
\begin{equation}
    x_t = \sum_{i=1}^D a_i x_{t-i}
\end{equation}
to $x$, obtaining residuals of variance $\sigma^2_x$. Then, we fit the model

\begin{equation}
    x_t = \sum_{i=1}^D a_i x_{t-i} + \sum_{i=1}^D b_i y_{t-i},
\end{equation}

obtaining residuals of variance $\sigma^2_{xy} < \sigma^2_{x}$.
Then, the prediction improvement is measured
by the Granger Causality estimator

\begin{equation}
GC =\frac{\sigma^2_x - \sigma_{xy}^2}{\sigma^2_x}
\end{equation}

The order of the autoregressive process, $D$, was decided using the Schwartz criterion \cite{schwarz1978estimating}.  In the dataset under study, the typical $D$ was the order of few days, with $80\%$ of the series having $D\le3$.
The GC significance was evaluated using 200 time-shift surrogates. The same procedure was used to assess the significance of all the other measures.
Finally, TE measures the information flow from one time series to another,
\begin{equation}
TE_{y\rightarrow x} = MI\left(y_t, x_{t-1:t-M} \big| y_{t-1:t-M}\right),
\end{equation}
that is the mutual information between $x$ and the past of $x$, conditional to the past of $y$. $M$ in this case is the order of the Takens embedding of $x$ and $y$. We estimated the $TE$ using a $k$-neighbours algorithm \cite{zhu2015contribution}, while $M$ was chosen again using the Schwartz criterion. Also in this case we set $k=4$.

Summarizing, for each topic, $n$, from the provincial agendas (raw time series), we calculated the $6\times 6$ symmetric matrices ${CC^n}_{i,j}$ and ${MI^n}_{i,j}$, and the $6\times 6$ matrices ${GC^n}_{i,j}$ and ${TE^n}_{i,j}$, which do not have a particular symmetry.

\section{Results}\label{Resultados}

Next, we present the results of the analysis of the agendas of the 20 topics. By inspecting the time series shown in Fig.~ \ref{fig_global_agenda}, it is clear that when a particular event trigger attention to a topic (for instance, the arrival of the Iranian--Venezuelan plane), the agenda presents a steep rise of attention that then decays. As expected, these topics are associated to one or a very few number of events and the attention vanishes after the events (as discussed before, the small noisy values before, and several days after each event can be considered numerical artifacts). On the other hand, topics that are frequently discussed in the media do not show very pronounced peaks. These “background” or “general” topics have a higher number of events due to their constant attention over time. 

\begin{figure}[tbp!]
    \centering
    \includegraphics[width=0.6\textwidth]{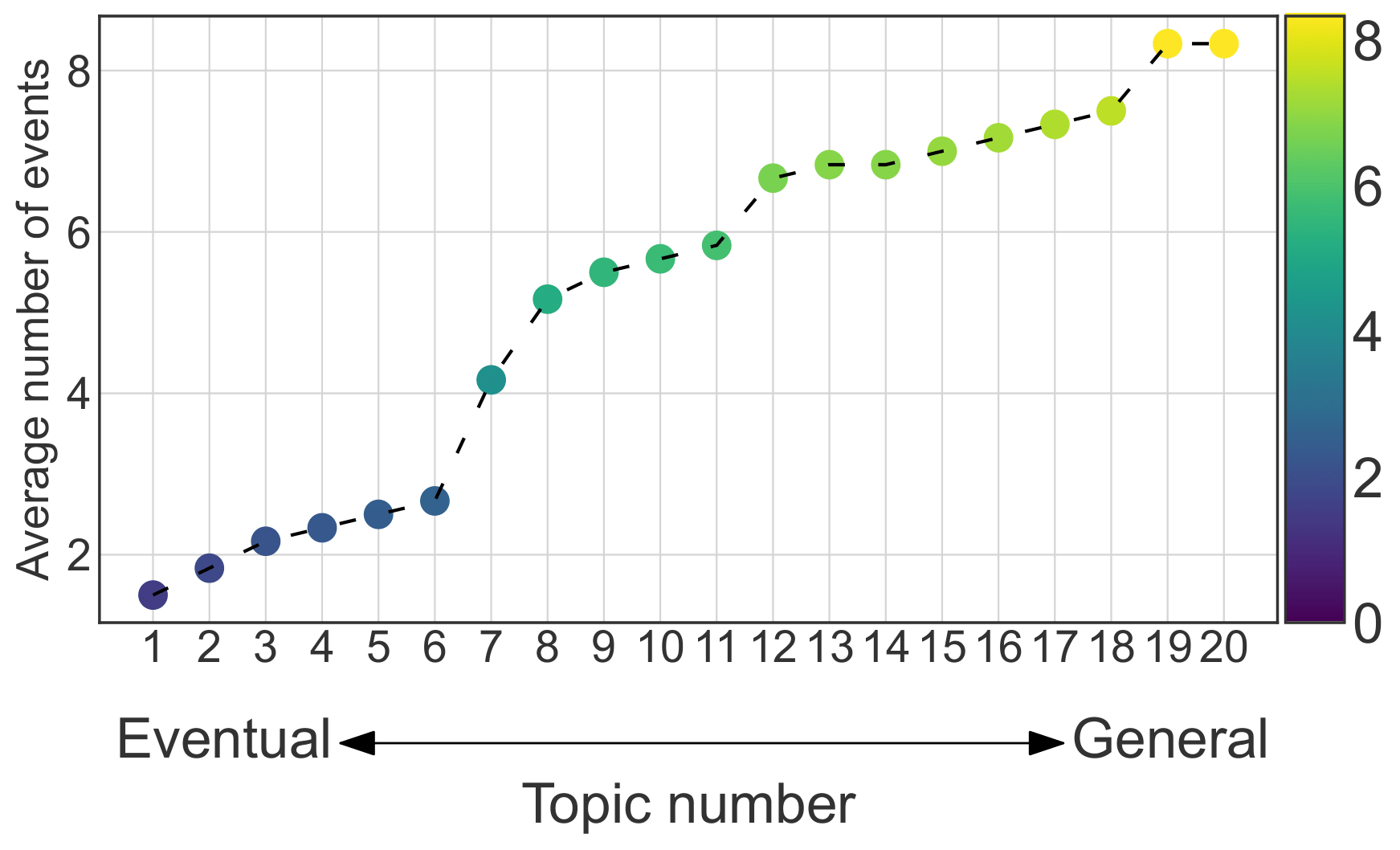}
    \caption{\label{fig_gradiente} \textbf{Differentiation of “eventual” and “general” topics.} Average number of events per province of each topic. The color gradient indicates which topics are more “Eventual” (dark blue) and which ones are more “General” (bright yellow).}
\end{figure}

The difference between “eventual” type and “general” type of topic is visualized in Fig. \ref{fig_gradiente} that displays the average number of detected events per province. The topics with few events (in the left part of the figure) are encapsulated by very specific words and are the most “eventual” topics, while the topics with the largest number of events (in the right part of the figure) are those that are encapsulated by less specific words and associated with more “general” themes.

\begin{figure}[tbp!]
    \centering
    \includegraphics[width=0.9\textwidth]{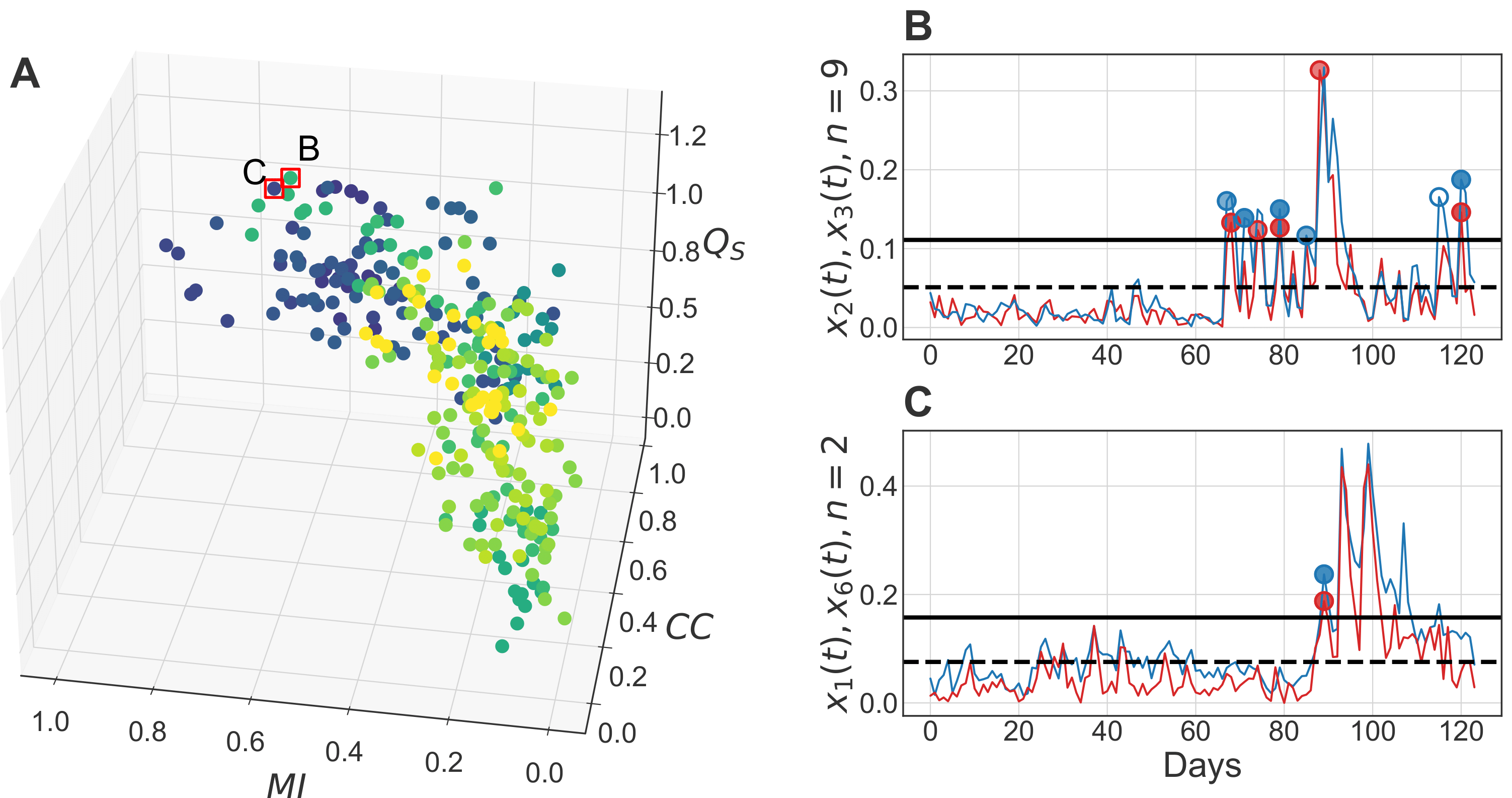}
    \caption{\label{fig_sym_sel} \textbf{Analysis of the symmetric quantifiers}. Panel \textbf{A} displays the 3D plot of ${Q^n_s}_{i,j}$, ${CC^n}_{i,j}$ and ${MI^n}_{i,j}$ for all topics and all pairs of provinces. The color indicates the topic, as in Fig. \ref{fig_gradiente}. The dots marked \textbf{B} and \textbf{C} have the maximum value of the sum of squares of the three metrics. Panels \textbf{B} and \textbf{C} display the corresponding time series: the agendas of topic “Vice President Trials” in Córdoba and Tucumán, and of topic “Vice President” in Buenos Aires and Santiago del Estero.}
\end{figure}

As explained in the previous section, for each topic we calculated the symmetric $6\times 6$ matrices ${Q^n_s}_{i,j}$, ${CC^n}_{i,j}$ and ${MI^n}_{i,j}$.
In Fig.~\ref{fig_sym_sel}\textbf{A} we plot all the values (for all the topics and for all pairs of provinces). In this figure the color code indicates the topic number and is the same as in Fig.~\ref{fig_gradiente}: dark blue (yellow) indicates topics with low (with large) average number of events, i.e., eventual (general) topics. It can be seen that eventual topics tend to have large values of ${Q_s}$, $CC$, and $MI$, while eventual topics have lower values.

The two pairs of time series with the largest synchronization level, identified as two pairs that have the largest values of $Q_s^2 +CC^2 + MI^2$, are shown in Figs. ~\ref{fig_sym_sel}\textbf{B} and \ref{fig_sym_sel}\textbf{C}. They correspond to the topics “Vice President Trials” and “Vice President”. The first one is for provinces Córdoba and Tucumán, while the second one, for Buenos Aires and Santiago del Estero. 

The synchronization between provinces' agendas can be visualized by using a network representation. To define the network, we selected the pairs of agenda time series that have largest similarity, measured in terms of the largest values of the sum $Q_s^2 +CC^2 + MI^2$. We chose this criterion because the three measures can be expected to provide complementary information, in the sense that they quantify different aspects in which two time series can be statistically similar ($CC$ detects linear correlations, $MI$ detects nonlinear correlations, and $Q_s$ detects synchronicity in the events defined in the time series).

Because these matrices are symmetric, we have $6 \times 5/2$ pairs of provinces and 20 topics, therefore, 300 possible pairs. We selected the top $10\%$ (30 pairs) with largest similarity. We remark that by selecting the largest values, we also select the most significant ones, because, the larger the $CC$ or $MI$ value is, the lower the p-value is (as shown in Fig.~\ref{fig_pvalues} in the Appendix).

\begin{figure}[tbp!] %% fig 7
    \centering
    \includegraphics[width=0.9\textwidth]{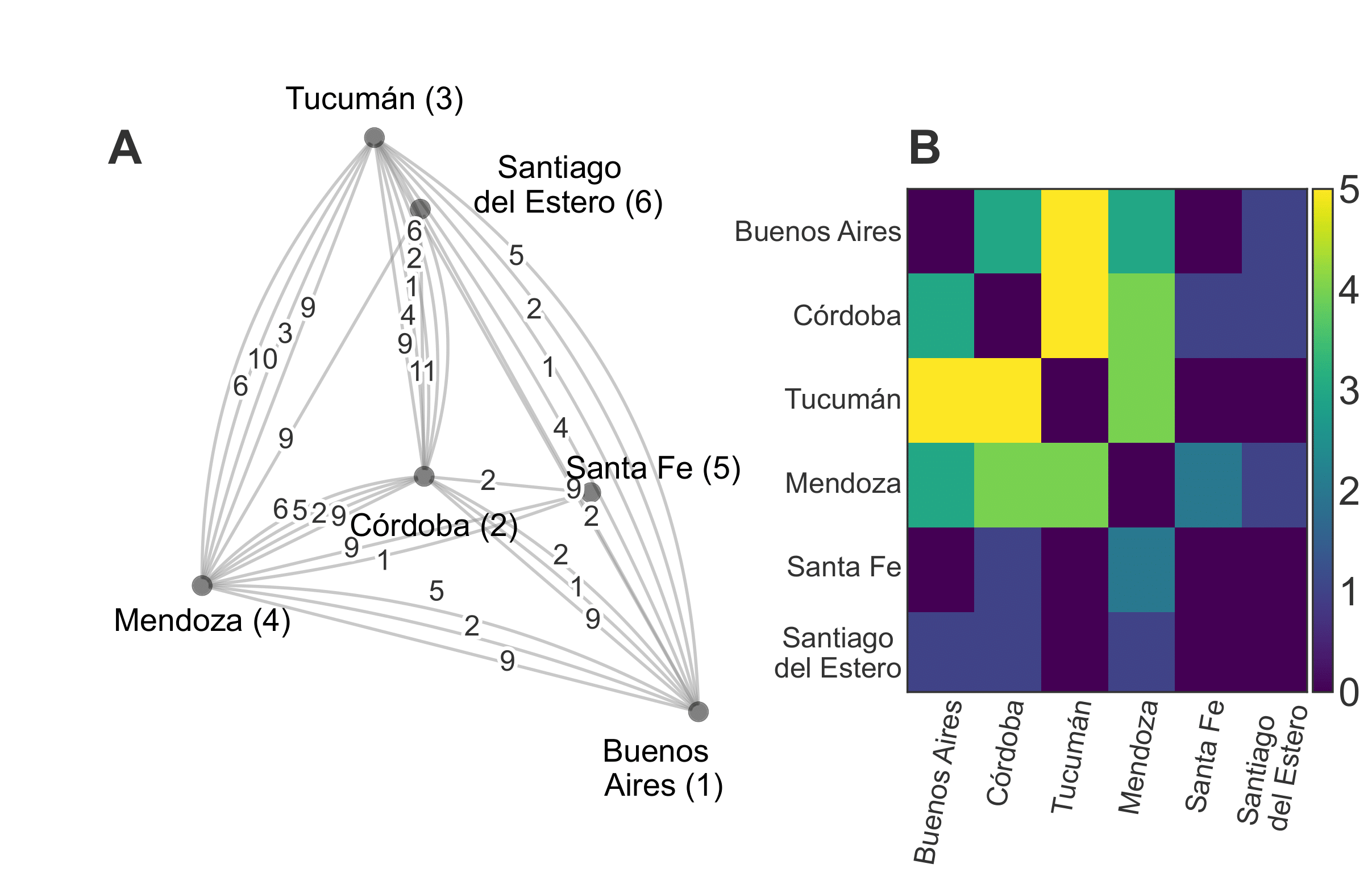}
    \caption{\label{fig_sync_net} \textbf{Network representation of the similarities uncovered between pairs of provincial agendas}.  \textbf{A} Network formed by the 30 links (top $10\%$) that have the largest values of the sum $Q_s^2 +CC^2 + MI^2$. The labels on the links indicate the topic.  \textbf{B} Weighted adjacency matrix where the weights (in color code) represent the number of (non-directed) links that are shared between pairs of provinces. }
\end{figure}

Figure \ref{fig_sync_net}\textbf{A} displays the network obtained in this way. The number of links that correspond to eventual topics (that have less than 4 events per province --20 links) doubles the number of links that correspond to general topics (that have 4 or more events events per province --10 links), showing again that the most synchronized links are those that correspond to topics with few and well-defined events.  {In Figure  \ref{fig_sym_timeseries} we show the time series corresponding to the top ten links with the largest similarity.}

Figure \ref{fig_sync_net}\textbf{B} displays the adjacency matrix of the same network without distinguishing topics.Here the color code indicates the number of links between pairs of provinces (i.e., the number of topics for with the temporal evolution of the agenda in the two provinces is very similar). From this matrix one could try to establish if there is a province that serves as a news broadcaster; however, no province has a clear news dissemination role. However, we find two pairs of provinces, Tucuman and Buenos Aires, and Tucuman and Cordoba, that share a large number of links, which indicates that the local agendas in various topics follow quite similar temporal evolution (an example was shown in Fig.~\ref{fig_sym_sel}\textbf{B} that displays the agendas of topic “Vice President Trials” in Tucuman and Cordoba).

As a final step, we study the direction of information diffusion by analyzing the matrices ${Q^n_a}_{i,j}$, ${GC^n}_{i,j}$ and ${TE^n}_{i,j}$.

\begin{figure}[tbp] %% fig 8
    \centering
    \includegraphics[width=0.85\textwidth]{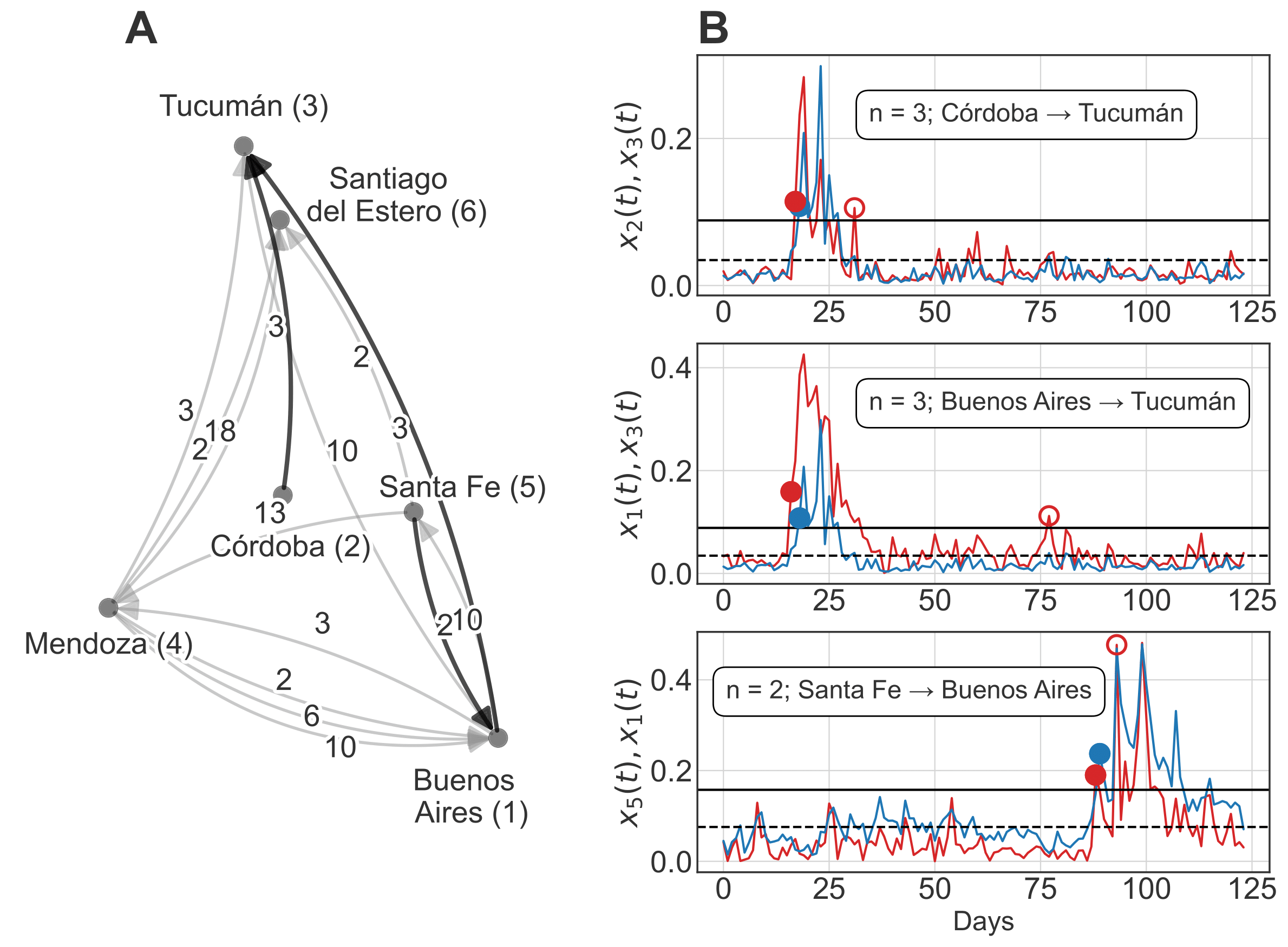}
    \caption{\label{fig_async_net_qa} \textbf{Analysis of directed links identified from event-based analysis.} \textbf{A} Directed network constructed by selecting the links that satisfy the conditions $ \left | Q_{a} \right |>0.6$ and $Q_{s} > 0.6$, which gives 14 directed links. \textbf{B} Time series corresponding to the three links marked with different line styles (the time series associated to all the links are shown in Fig.~\ref{fig_sm_qa}). {In each panel we can observe that the event of the red time series precedes the event of the blue one with one day of difference.}} 
\end{figure}

The absolute values of the elements of $Q_{a}$ provide information about the level of asymmetry of the synchronization between two provinces, and their sign indicate which province is the leader, and which one lags behind (low $|Q_{a}|$ values indicate that the events in the two provinces alternate, while high $|Q_{a}|$ values indicate that events in one province tend to precede the events in the other province).

To represent the most relevant links in a directed network, we selected pairs of provinces whose values of $ \left | Q_{a} \right |$ and $Q_{s}$ are both larger than $0.6$. The network obtained in this way, shown in Fig. \ref{fig_async_net_qa}\textbf{A} has 14 links. In Fig.~\ref{fig_async_net_qa}\textbf{B}, the time series of the three links marked with different line styles (dash, dotted, dash-dotted) are shown. {Here, we can observe that the event of the red time series precedes the event of the blue one with one day of difference}. It is important to remark that the network obtained depends on the threshold used to select the largest $|Q_a|$ and $Q_s$ values (and also, on the thresholds used to detect the events). We remark that, while the links that are selected vary with these thresholds, the topics that are associated to these links are always the most eventual ones.

\begin{figure}[tbp] %% fig 9
    \centering
    \includegraphics[width=0.85\textwidth]{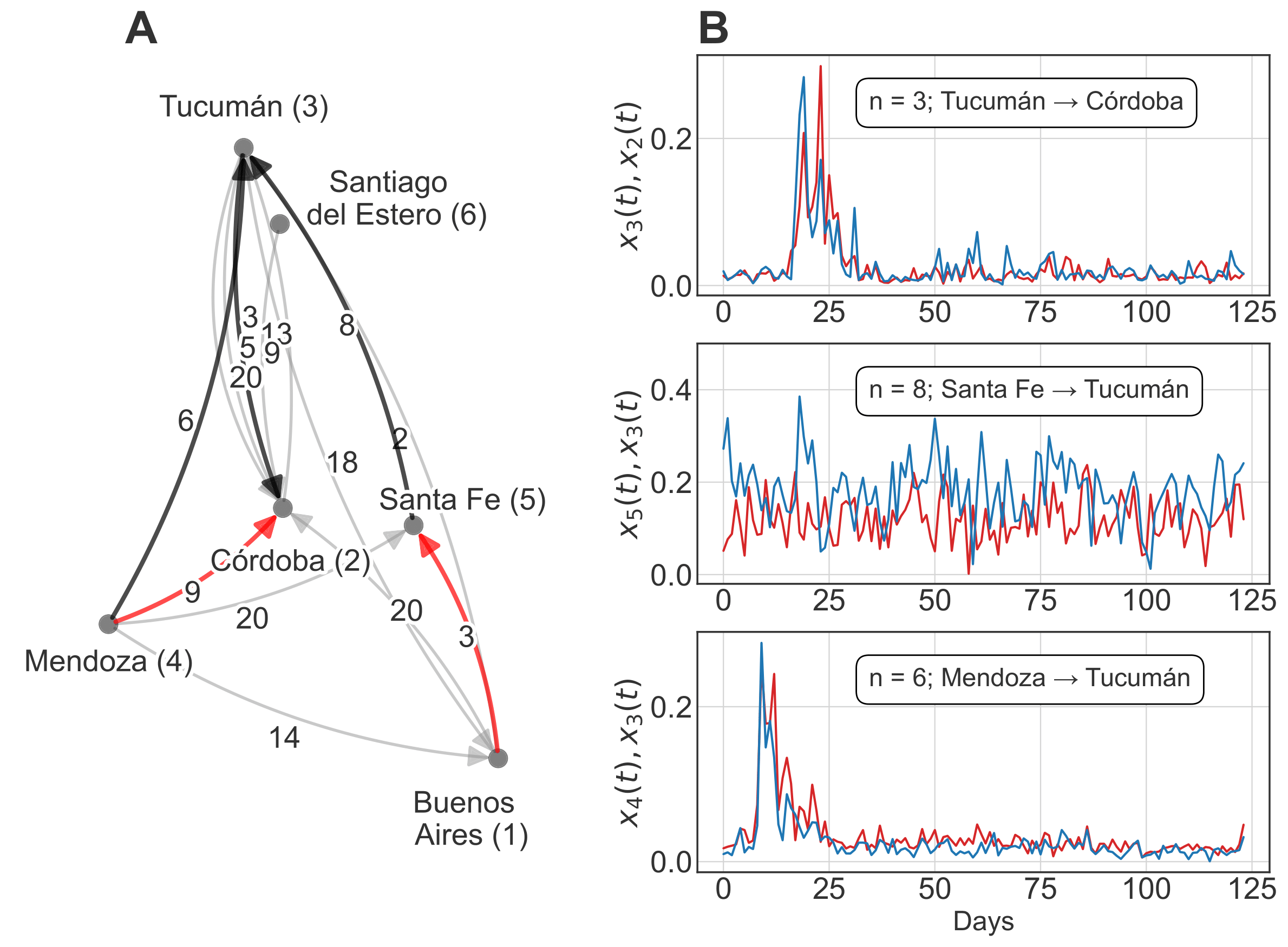}
    \caption{\label{fig_async_net_te} \textbf{Analysis of directed links identified from transfer entropy analysis.} \textbf{A} Directed network constructed by selecting the 14 links (the same number of links as in the network shown in Fig.~\ref{fig_async_net_qa}) with highest $TE$ values. {The black edges are the ones presented in \textbf{B}, while the red edges are the ones that match with the $GC$ network}. \textbf{B} Time series corresponding to the three links marked with different colors (the time series associated to all the links are shown in Fig.~\ref{fig_sm_te}). {The upper panel shows the series for Tucumán and Córdoba where the $TE$ detected an inverse direction of the flow compared to the $Q_{a}$ method.}}
\end{figure}

\begin{figure}[tbp] %%% fig 10
    \centering
    \includegraphics[width=0.85\textwidth]{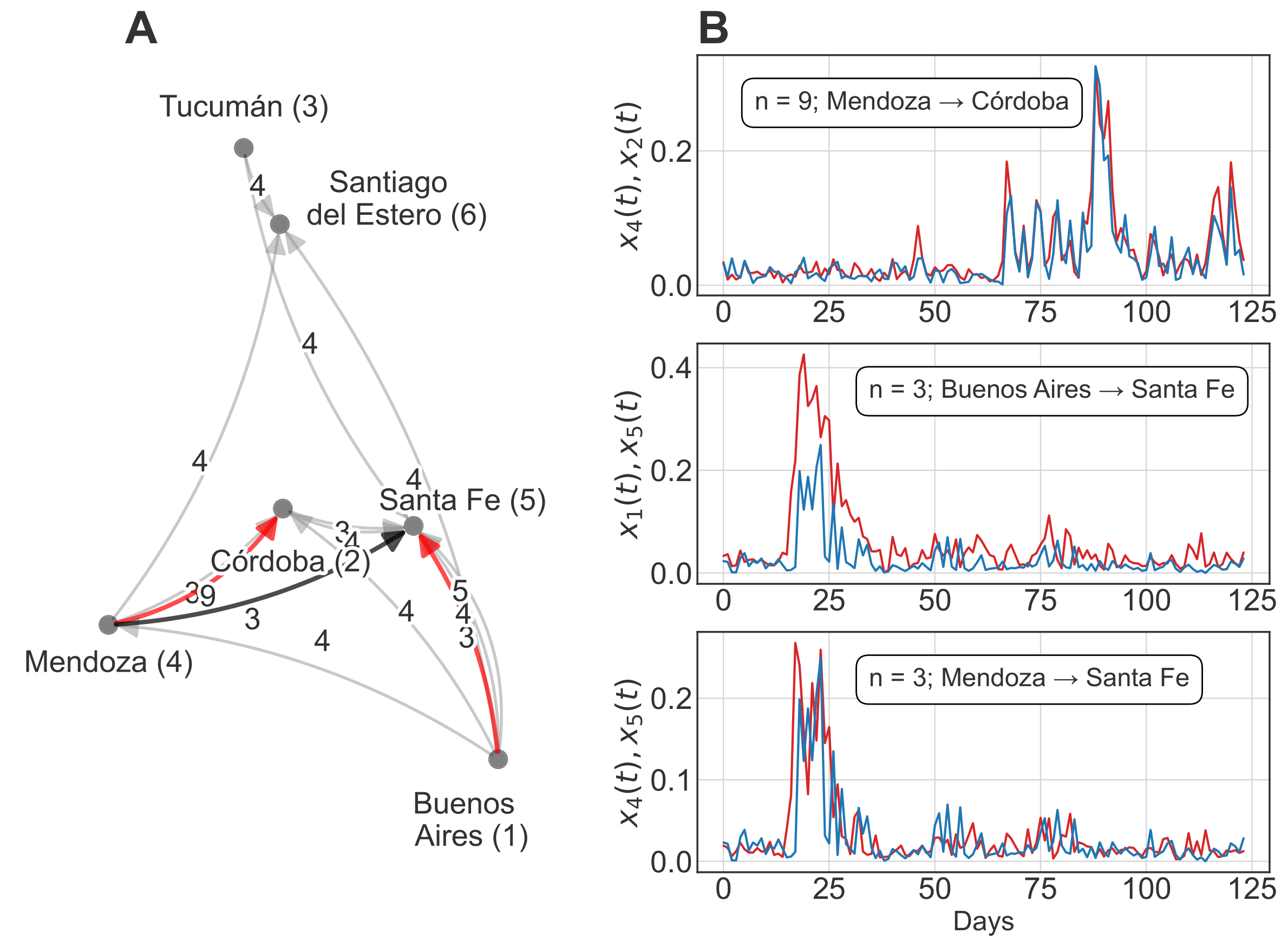}
    \caption{\label{fig_async_net_gc} \textbf{Analysis of directed links identified from Granger causality analysis}. \textbf{A} Directed network with the 14 links that have highest $GC$ values. {The time series corresponding to coloured edges are displayed in panel \textbf{B}, the red edges [for Mendoza and Córdoba for the topic “Vice President Trials” and for Buenos Aires and Santa Fé for the topic “Diplomatic Conflict”] are also present in the network shown in Fig. \ref{fig_async_net_te}\textbf{A}.}}
\end{figure}

The networks obtained from $TE$ and $GC$ are displayed in Figs.~\ref{fig_async_net_te} and \ref{fig_async_net_gc} respectively. In both cases, we selected the 14 links with largest value, to make them comparable with the previous case. The $TE$ network doesn't share any links with the $Q_{a}$ network. Even though the two methods detect an interaction between Córdoba and Tucumán for the topic “Diplomatic Conflict”, in the  $Q_{a}$ network it goes from Córdoba to Tucumán (as can be observed in the temporal series of top panel of Figure \ref{fig_async_net_qa}\textbf{B}), while the $TE$ method sets the inverse relation. When comparing the $TE$ and $GC$ networks, they only share two links (colored in red in both figures): between Buenos Aires and Santa Fé for topic 3 (Diplomatic Conflict) and between Mendoza and Córdoba for topic 9 (Vice President Trials). 

Moreover, the two networks do not share any link with the network shown in Fig.~\ref{fig_async_net_qa}, obtained by thresholding $Q_s$ and $|Q_{a}|$. We believe that the discrepancy can be due to three main reasons: First, the event-based approach filters out small noisy fluctuations (which, as discussed before, can be an artifact), but $GC$ and $TE$ measures do not because they are computed from the raw time series. Secondly, the time series are strongly non-stationary, and one of the implicit assumption of both $GC$ and $TE$ is that the time series under study are stationary. The quantities $Q_s$ and $Q_{a}$, instead, do not have this kind of restriction. Thirdly, $Q_s$ and $Q_a$ are symmetric and asymmetric measures respectively, thus capture the strength and the net direction of information flow between two provinces, whereas $GC$ and $TE$ are not asymmetric measures and do not return a net direction of information flow. 

While asymmetric quantities could have been  defined by using the differences $GC^n_{i,j} - GC^n_{j,i}$ and $TE^n_{i,j} - TE^n_{j,i}$, this easy solution has the drawback that the pairs of values ($GC^n_{i,j}, GC^n_{j,i}$) and ($TE^n_{i,j}, TE^n_{j,i}$) may not have a similar level of significance (i.e., they may  have very different p-values). Therefore the largest values of $|GC^n_{i,j} - GC^n_{j,i}|$ and $|TE^n_{i,j} - TE^n_{j,i}|$ may not capture the most significant links. 

\section{Conclusion and discussion}\label{Conclusiones}

We have analyzed data collected from Argentinian newspapers published in six main provinces during a period of four months. We classified the articles into 20 non-orthogonal topics, and obtained a set of 120 time series representing the evolution of the media agenda in the provinces. To capture statistical similarities and dependencies, the time series were analyzed using six bivariate quantifiers, which can be classified into two main types: those computed from the raw time series (the cross-correlation, the mutual information, Granger causality, and transfer entropy) and those computed from events defined in the time series (symmetric and asymmetric measures, $Q_{s}$ and $Q_{a}$, respectively). 

To define events in the time series, we used two thresholds, one to identify when the media attention picks up and another to detect when the attention decays and another event may take place latter.

By analyzing the number of events for the different topics in the different provinces, we were able to qualitatively distinguish them into two categories (Fig \ref{fig_gradiente}): “General” topics, which are continuously active in the press agenda and typically display several events in each province, and “Eventual” topics, which are topics that have attracted the media attention at particular times, and often show only one or few events in each province. Examples of eventual topics include “Assault on Vice President” (topic number 1), “Vice President” (topic number 2)  and “Diplomatic conflict” (topic number 3), while examples of generalist topic include “Opposition” (topic number 18), “International Affairs” (topic number 19) and “Social Assistance” (topic number 20).

The six bivariate measures used to analyze pairs of times series for a given topic in different provinces can also be classified into two types: undirected ($CC$, $MI$, and $Q_{s}$) and directed ($GC$, $TE$, and $Q_{a}$). We used the undirected measures to identify statistical similarities between pairs of time series and the directed ones to determine the direction of information flow. To analyze the performance of these measures in detecting meaningful connections and the main paths of information diffusion, we tested different strategies to identify, for each topic, the most significant links between towns.

Because the three undirected measures provide complementary information, in the sense that they quantify different aspects in which two time series can be statistically similar, we selected the links whose sum, $CC^2+MI^2+Q_{s}^2$, was higher than a given threshold. The network obtained, shown in Fig. \ref{fig_sync_net}, certainly depends on the threshold used, however, an inspection of the time series of the links identified in this way, revealed clear similarities in the temporal variation of the media attention (two examples are shown in Fig. \ref{fig_sym_sel}\textbf{B}).

The three directed measures we used, $GC$, $TE$, and $Q_{a}$, can also be expected to provide complementary information. $GC$ and $TE$ are computed from raw data, while $Q_{a}$ results from the relative timing of media attention events. We found that $GC$ and $TE$ often gave contradicting results, with the dependency returned by $GC$ being the opposite of that returned by $TE$. In this sense, we found that thresholding $GC$ and $TE$ was not useful for identifying clear dependencies in the process of information dissemination, at least with the hyper-parameters used in this study. Thus, we used a simple pruning strategy to identify directed links: we selected links whose $Q_{s}$ and $|Q_{a}|$ were both above a threshold. 

Again, the network obtained depended on the threshold used, and an inspection of the time series corresponding to the links identified revealed situations in which the attention on a given topic in two towns began simultaneously or almost simultaneously. As it can be seen in Fig. \ref{fig_async_net_qa}, using a threshold of 0.6 allowed us to identify some topics and provinces where the attention started first, followed by the growth of attention in other provinces. 

It is important to remark that the event-based measures, $Q_{s}$ and $Q_{a}$, have, besides the thresholds used to define events, another hyper-parameter that plays an important role, which is the time interval allowed between two events that occur in the same topic in different provinces, to count them as related. In general, increasing $\tau$ results in more values of $Q_{a}$ and $Q_{s}$ that are above the thresholds. This may be because, when the attention on topic $n$ begins in province $i$, in province $j$ the attention is on a different topic, and we need to allow for more than one day to be able to detect the fact that in province $j$, topic $n$ gains attention. Of course, using values of $\tau$ that are too long leads to count as synchronous events that can be unrelated.

Taken together, our results show that the process of information diffusion in newspapers in Argentina is very fast, and genuinely related “events” in different provinces are most often simultaneous. It is worth noticing that, in the networks found using the different thresholding strategies, no particular province was found to play a clear role of “information hub”. The quasi-simultaneous activation of a topic in different provinces can be understood because all of them are triggered by the same events and looks almost synchronous within a one day resolution time. However, information spreading on social media or large press agencies can also have an effect  acting as global drivers.

For future work, it will be interesting to test the generality of our conclusions, by analyzing data extracted from articles in newspapers in other countries. It will be also interesting to correlate media attention on socio-economic topics with financial time series that effectively capture the temporal evolution of the economy, at the provincial level, national level, and global international level.

\begin{acknowledgments}
C.M. acknowledges the support of ICREA ACADEMIA, AGAUR (2021 SGR 00606), and Ministerio de Ciencia e Innovación Spain (Project No. PID2021-123994NB-C21). P.B acknowledges the support of the Universidad de Buenos Aires (UBA)  through Grant UBACyT, 20020220100181BA and the Agencia Nacional de Promoción de la Investigación, el Desarrollo Tecnológico y la Innovación through Grant No. PICT-2020-SERIEA-00966.
\end{acknowledgments}

\section*{Data availability}
The time series analysed, the code to calculate the events, coincidences and the matrices are available at \url{https://github.com/luciolgarcia/News_diffution}

\bibliography{Main}

\appendix
\section{}\label{secA1}

\begin{figure}[htb]
    \centering
    \includegraphics[width=0.95\textwidth]{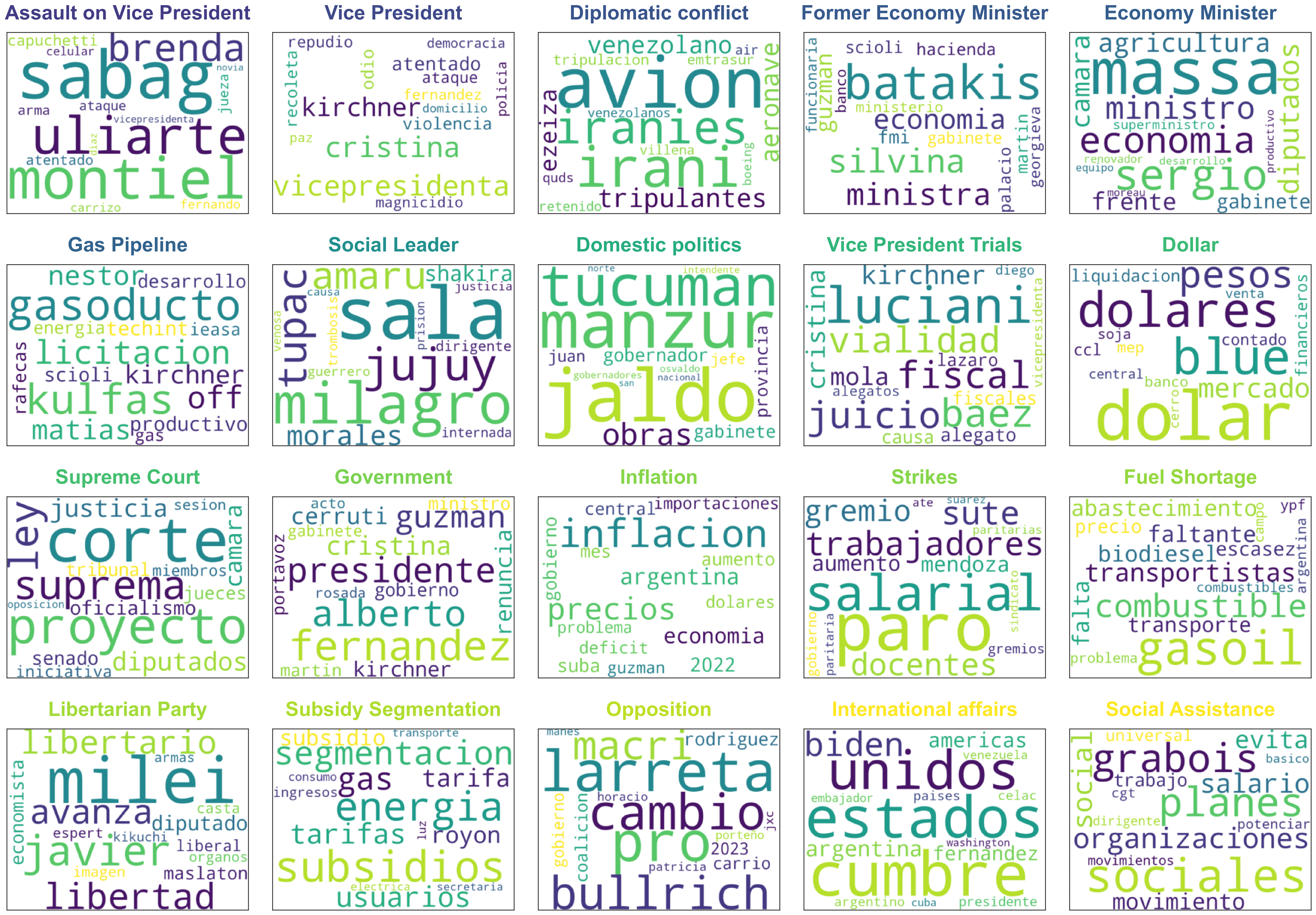}
    \caption{\label{fig_wordclouds} Wordclouds of the 20 topics}
\end{figure}

\begin{table}[htbp]
\centering
\caption{\label{tab:topics} Topics labels assigned and their most relevant five keywords which are used to describe it.}
\resizebox{\textwidth}{!}{\begin{tabular}{|c|l|l|}
\hline
\textbf{Number} & \textbf{Topic label} & \textbf{Keywords} \\ \hline
1 & Assault on Vice President & sabag, montiel, uliarte, brenda, capuchetti
\\ \hline
2 & Vice President & vicepresidenta, cristina, kirchner, atentado, odio
\\ \hline
3 & Diplomatic conflict & avion, irani, iranies, tripulantes, aeronave 
\\ \hline
4 & Former Economy Minister & batakis, silvina, ministra, economia, guzman
\\ \hline
5 & Economy Minister & massa, sergio, economia, ministro, diputados
\\ \hline
6 & Gas Pipeline & gasoducto, kulfas, licitacion, off, matias
\\ \hline
7 & Social Leader & sala, milagro, jujuy, tupac, amaru
\\ \hline
8 & Domestic politics & jaldo, manzur, tucuman, obras, gobernador
\\ \hline
9 & Vice President Trials & luciani, baez, fiscal, juicio, vialidad
\\ \hline
10 & Dollar & dolar, blue, dolares, pesos, mercado
\\ \hline
11 & Supreme Court & corte, proyecto, suprema, ley, diputados
\\ \hline
12 & Government & fernandez, alberto, presidente, guzman, cristina
\\ \hline
13 & Inflation & inflacion, precios, argentina, economia, 2022
\\ \hline
14 & Strikes & paro, salarial, trabajadores, sute, docentes
\\ \hline
15 & Fuel Shortage & gasoil, combustible, transportistas, faltante, abastecimiento
\\ \hline
16 & Libertarian Party & milei, javier, libertad, avanza, libertario
\\ \hline
17 & Subsidy Segmentation & subsidios, energia, segmentacion, gas, usuarios
\\ \hline
18 & Opposition & larreta, pro, cambio, bullrich, macri
\\ \hline
19 & International affairs & cumbre, estados, unidos, biden, argentina
\\ \hline
20 & Social Assistance & sociales, planes, grabois, organizaciones, social
\\ \hline
\end{tabular}}
\end{table}

\begin{figure}[htb]
    \centering
    \includegraphics[width=0.7\textwidth]{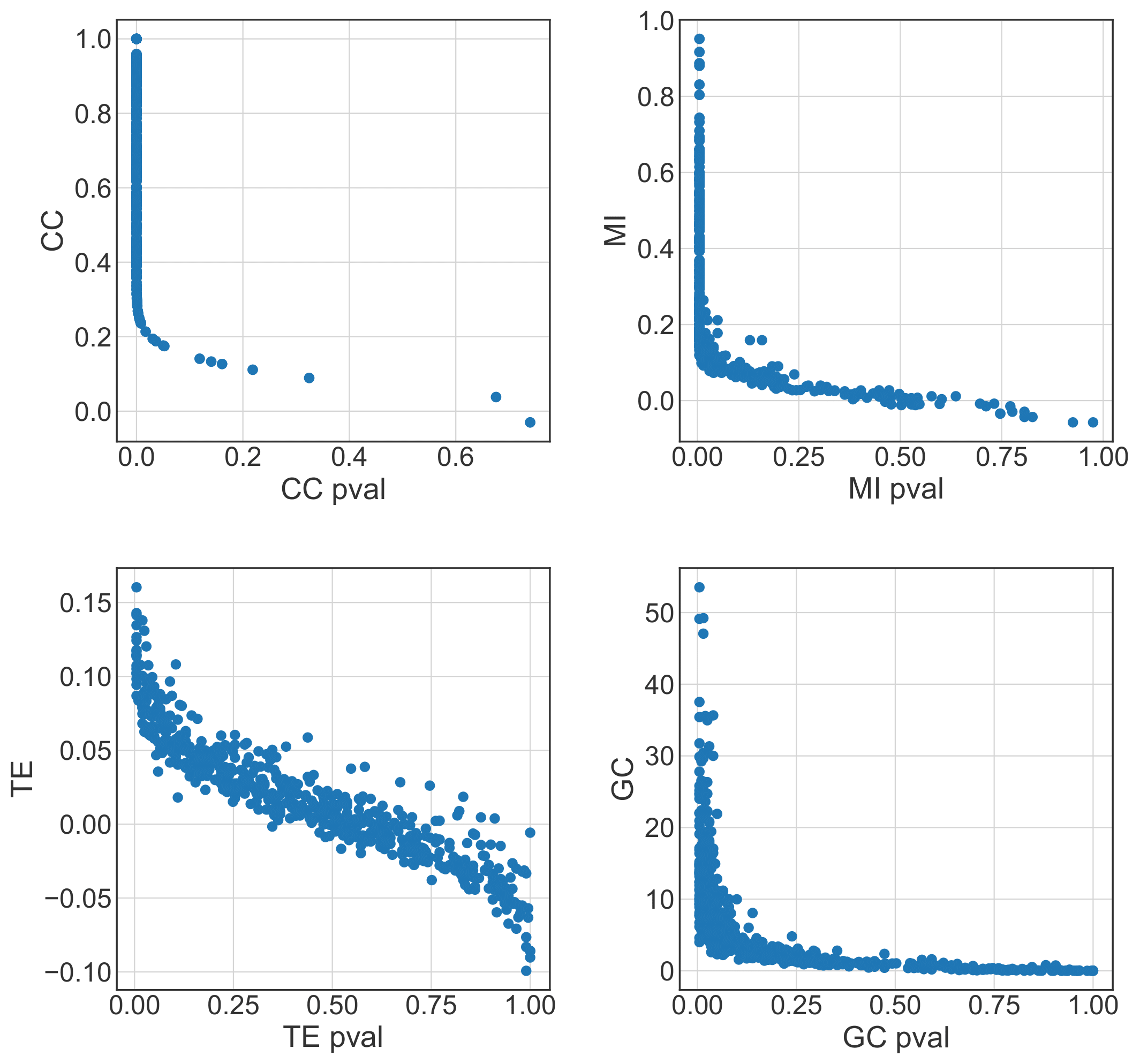}
    \caption{\label{fig_pvalues} Relationships between the metrics computed from the raw time series (cross-correlation, mutual information, Granger causality and transfer entropy) with their p-values. It can be seen that the lower p-values correspond to the largest values of the metrics. For this reason, selecting the largest values implies selecting the most significant ones.}
\end{figure}

\begin{figure}[htb]
    \centering
    \includegraphics[width=0.7\textwidth]{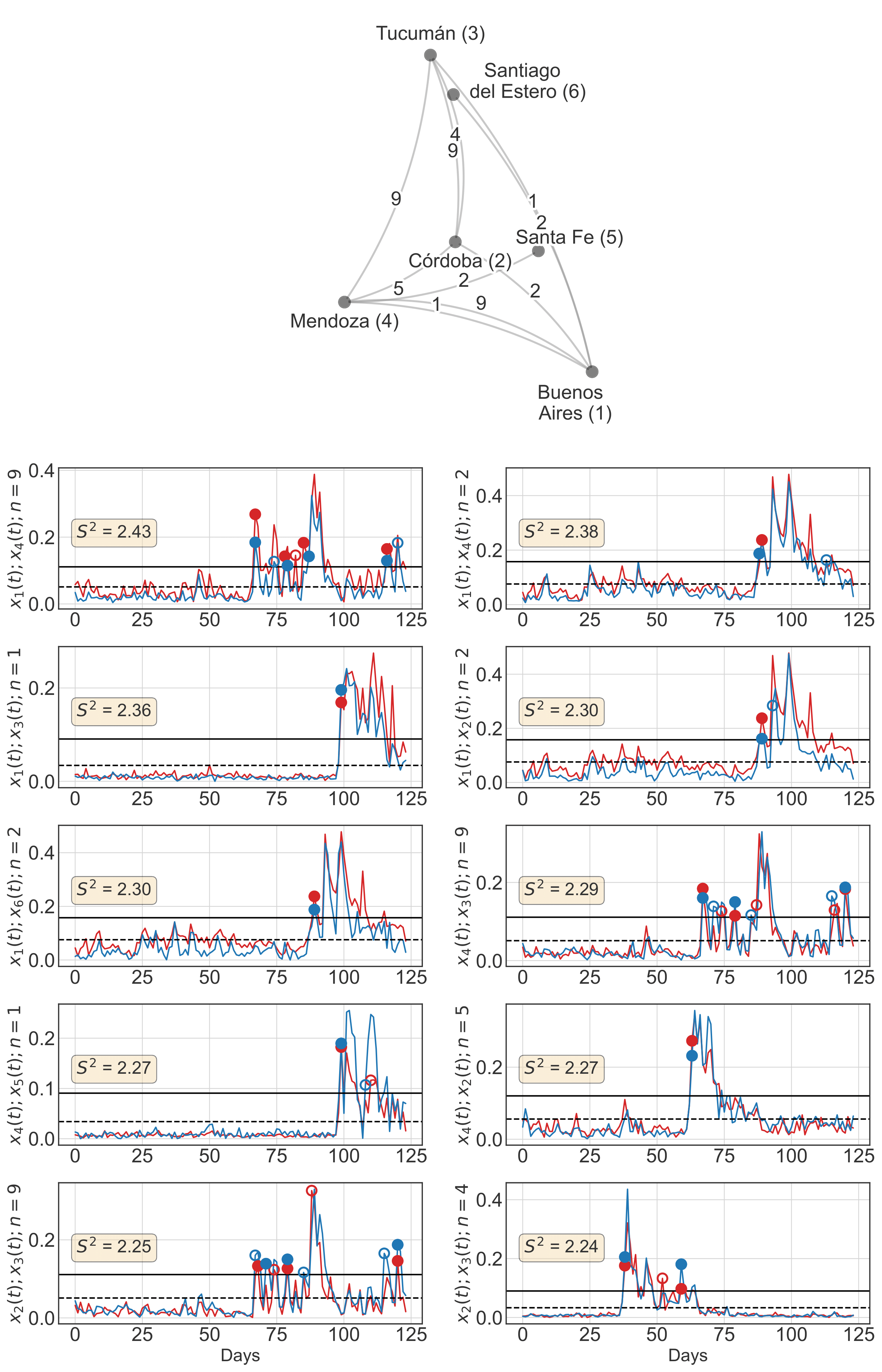}
    \caption{\label{fig_sym_timeseries} \textbf{Top panel}: Network constructed by selecting the 10 top links which maximizes the squared sum of the symmetric measures.
    \textbf{Bottom panels} display the two time series of each link.}
\end{figure}

\begin{figure}[htb]
    \centering
    \includegraphics[width=0.7\textwidth]{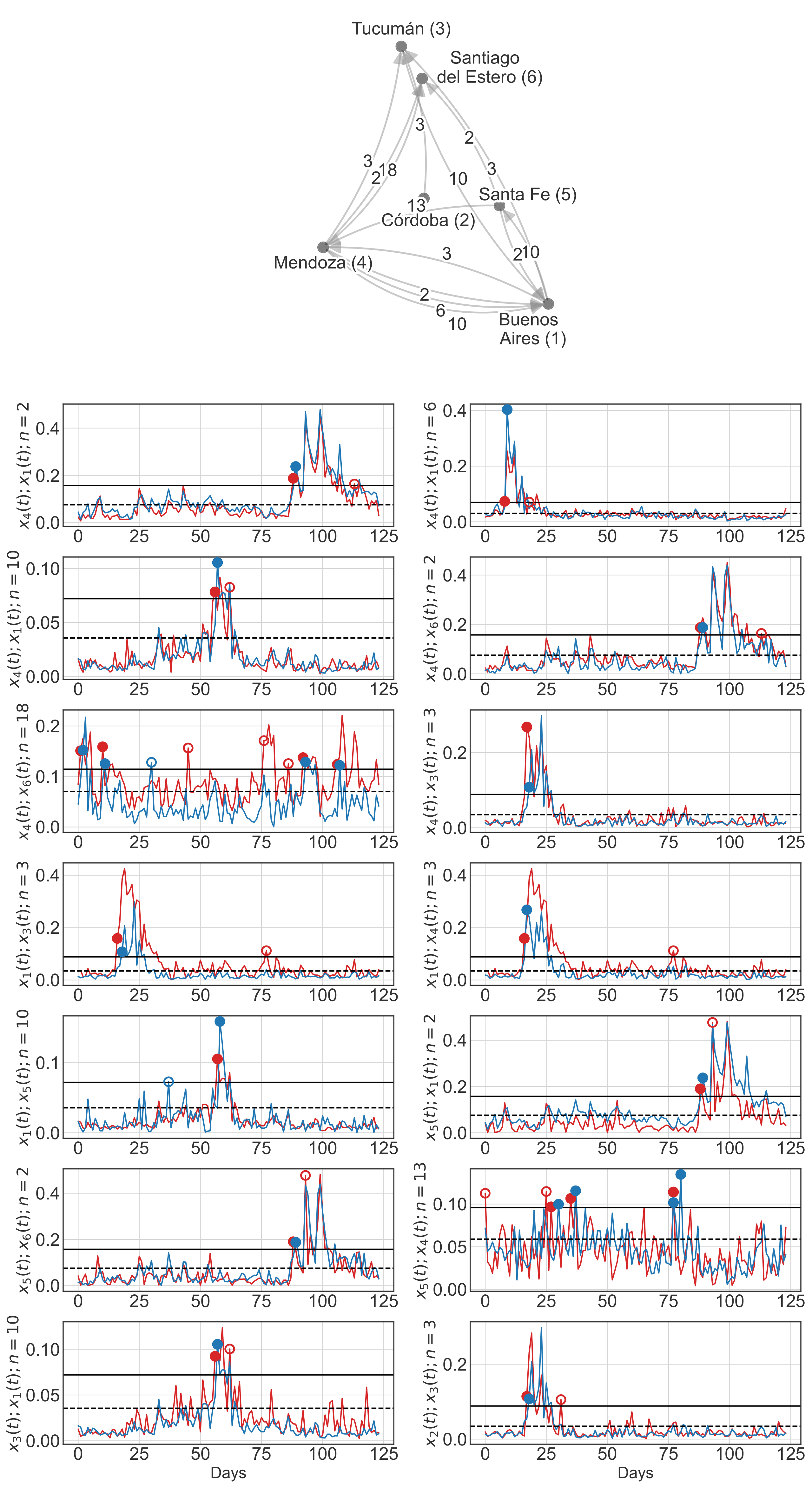}
    \caption{\label{fig_sm_qa} \textbf{Top panel}: Network constructed by selecting the links with $Q_{a}$ and $Q_{s} > 0.6$. This yields 14 directed links. \textbf{Bottom panels} display the two time series of each link.}
\end{figure}

\begin{figure}[htb]
    \centering
    \includegraphics[width=0.7\textwidth]{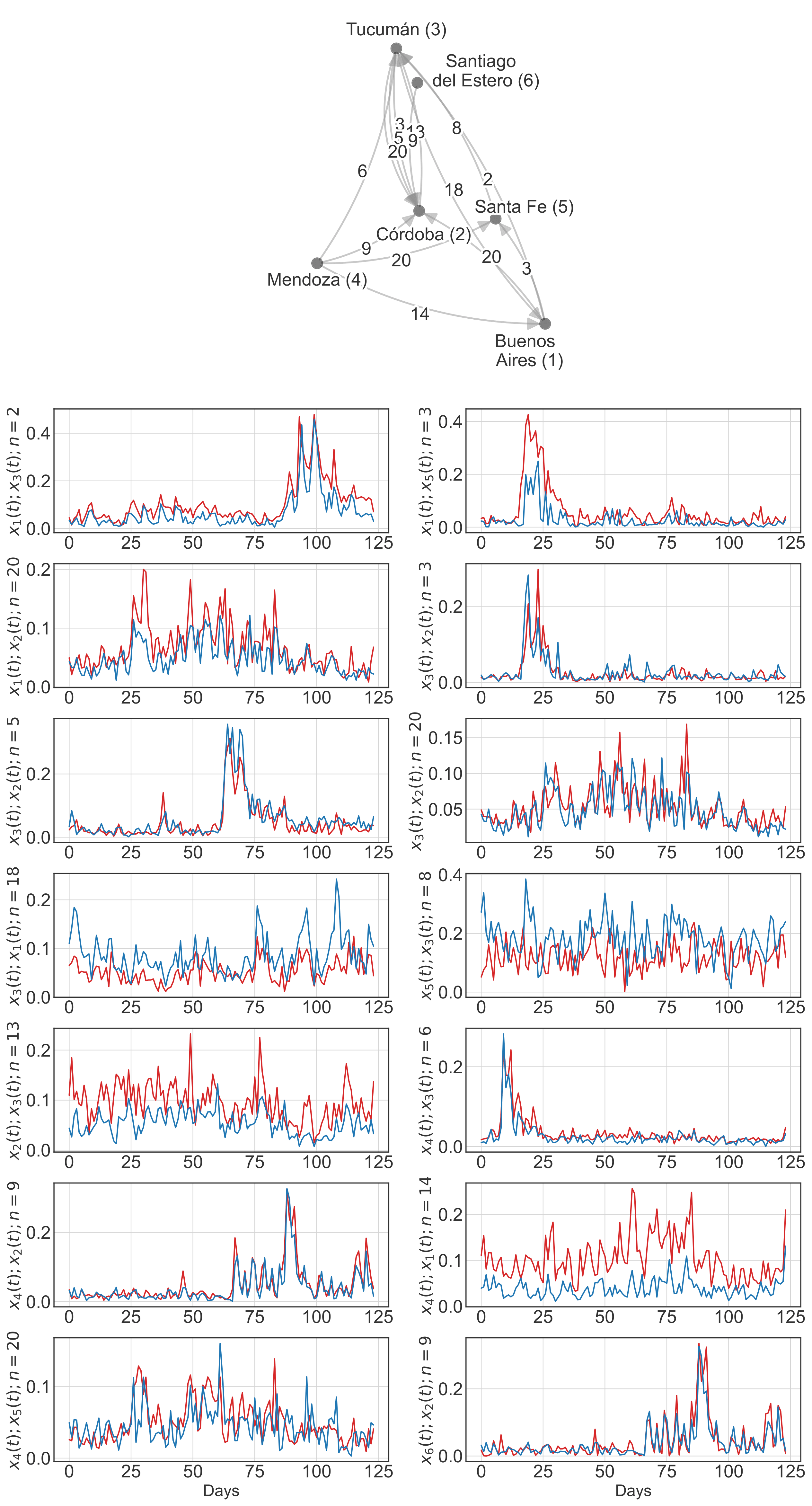}
    \caption{\label{fig_sm_te} \textbf{Top panel}: Network constructed by selecting the 14 links with highest $TE$ value. \textbf{Bottom panels} display the two time series of each link.}
\end{figure}

\begin{figure}[htb]
    \centering
    \includegraphics[width=0.7\textwidth]{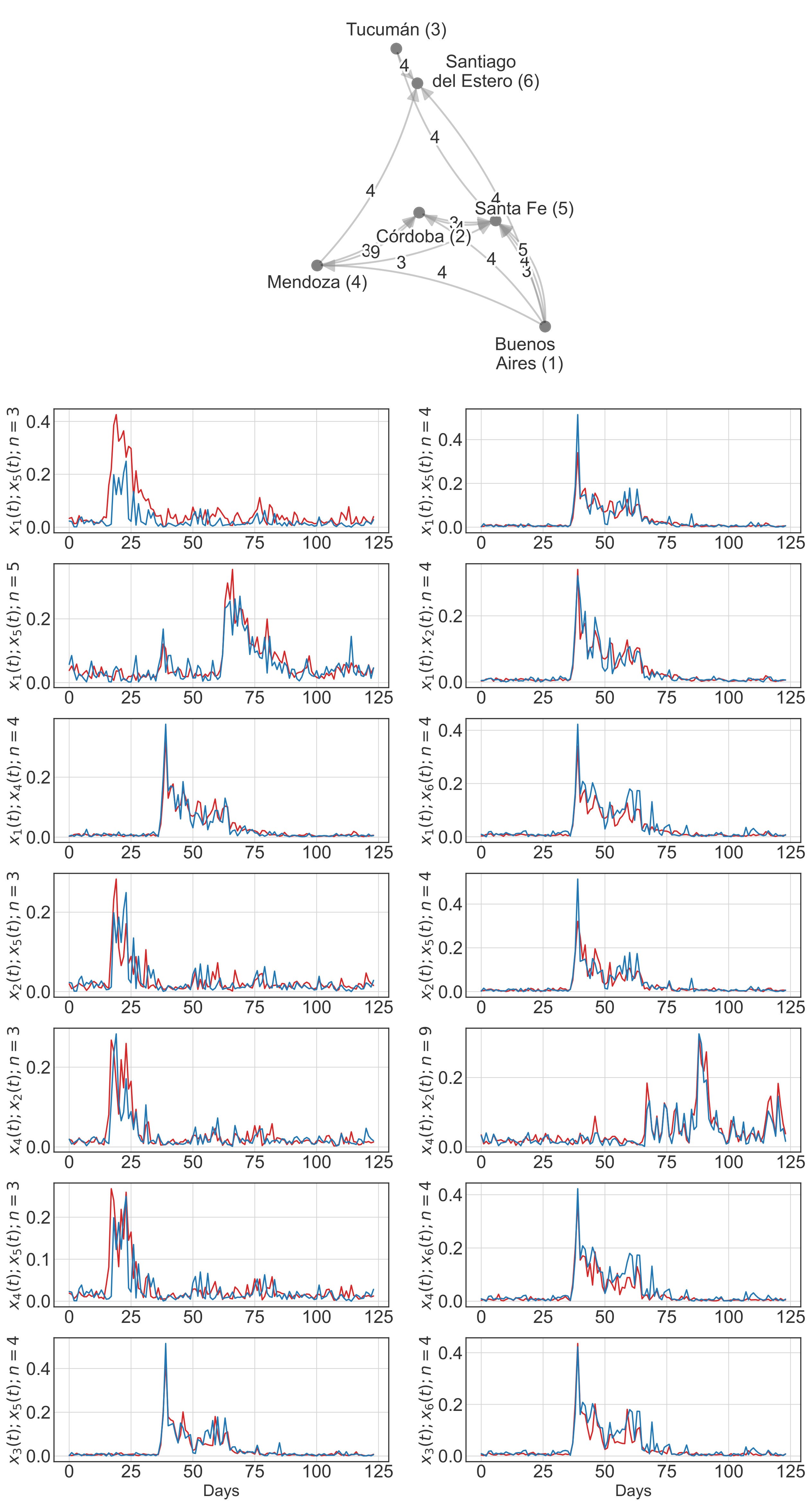}
    \caption{\label{fig_sm_gc} \textbf{Top panel}: Network constructed by selecting the 14 links with highest $GC$ value. \textbf{Bottom panels} display the two time series of each link.}
\end{figure}

\end{document}